\documentclass[journal]{IEEEtran}
\usepackage{graphicx}
\usepackage{subcaption}
\usepackage{blindtext}
\usepackage{etoolbox}
\usepackage{multirow}
\usepackage{array}
\usepackage[noadjust]{cite}
\usepackage{url}
\usepackage{diagbox}
\usepackage{mathtools}
\usepackage{amsmath}
\usepackage{amssymb}
\usepackage{empheq}
\usepackage{bm}
\usepackage{cases}
\usepackage{color}
\usepackage{gensymb}
\usepackage{algorithm}
\usepackage{algpseudocode}
\usepackage{amsthm}
\usepackage[many]{tcolorbox}
\usepackage[shortlabels]{enumitem}
\usepackage{booktabs}
\usepackage{empheq}
\usepackage{textgreek}

\usepackage[makeroom]{cancel}
\makeatletter

\def\cantox@vector#1#2#3#4#5#6#7#8{%
  \dimen@.5\p@
  \setbox\z@\vbox{\boxmaxdepth.5\p@
   \hbox{\kern-1.2\p@\kern#1\dimen@$#7{#8}\m@th$}}%
  \ifx\canto@fil\hidewidth  \wd\z@\z@ \else \kern-#6\unitlength \fi
  \ooalign{%
    \canto@fil$\m@th \CancelColor
    \vcenter{\hbox{\dimen@#6\unitlength \kern\dimen@
      \multiply\dimen@#4\divide\dimen@#3 \vrule\@depth\dimen@\@width\z@
      \vector(#3,-#4){#5}%
    }}_{\raise-#2\dimen@\copy\z@\kern-\scriptspace}$%
    \canto@fil \cr
    \hfil \box\@tempboxa \kern\wd\z@ \hfil \cr}}

\def\bcancelto#1#2{\let\canto@vector\cantox@vector\cancelto{#1}{#2}}
\makeatother

\makeatletter
\newcommand*\rel@kern[1]{\kern#1\dimexpr\macc@kerna}

\newcommand*\widebar[1]{%
  \begingroup
  \def\mathaccent##1##2{%
    \rel@kern{0.8}%
    \overline{\rel@kern{-0.8}\macc@nucleus\rel@kern{0.2}}%
    \rel@kern{-0.2}%
  }%
  \macc@depth\@ne
  \let\math@bgroup\@empty \let\math@egroup\macc@set@skewchar
  \mathsurround\z@ \frozen@everymath{\mathgroup\macc@group\relax}%
  \macc@set@skewchar\relax
  \let\mathaccentV\macc@nested@a
  \macc@nested@a\relax111{#1}%
  \endgroup
}
\makeatother

\makeatletter
\def\widebreve{\mathpalette\wide@breve}
\def\wide@breve#1#2{\sbox\z@{$#1#2$}%
     \mathop{\vbox{\m@th\ialign{##\crcr
\kern0.08em\brevefill#1{0.8\wd\z@}\crcr\noalign{\nointerlineskip}%
                    $\hss#1#2\hss$\crcr}}}\limits}
\def\brevefill#1#2{$\m@th\sbox\tw@{$#1($}%
  \hss\resizebox{#2}{\wd\tw@}{\rotatebox[origin=c]{90}{\upshape(}}\hss$}
\makeatother

\theoremstyle{definition}

\newtcolorbox{framefloat}[1][]{fonttitle=\bfseries, titlerule=0pt, boxrule=0.5pt,
  colframe=black,colback=white,float=!t,#1}
  
\newcounter{mytempeqncnt}

\makeatletter

\docsvlist{\@oddhead,\@evenhead,\ps@headings,\ps@IEEEtitlepagestyle,\ps@IEEEpeerreviewcoverpagestyle}
\makeatother

\makeatletter
\newcommand\fs@spaceruled{\def\@fs@cfont{\bfseries}\let\@fs@capt\floatc@ruled
  \def\@fs@pre{\vspace{1\baselineskip}\hrule height.8pt depth0pt \kern2pt}%
  \def\@fs@post{\kern2pt\hrule\relax}%
  \def\@fs@mid{\kern2pt\hrule\kern2pt}%
  \let\@fs@iftopcapt\iftrue}
\makeatother

\allowdisplaybreaks

\begin{document}

\title{Reconfigurable Antenna Arrays\\ With Tunable Loads: Expanding Solution Space\\ via Coupling Control}
\author{$\text{Elio Faddoul}$, \textit{Member, IEEE}, $\text{Konstantinos Ntougias}$, \textit{Member, IEEE}, and $\text{Ioannis Krikidis}$, \textit{Fellow, IEEE} \thanks{The authors are with the Department of Electrical and Computer Engineering, University of Cyprus, Nicosia, Cyprus (e-mail: \{efaddo01, kntoug01, krikidis\}@ucy.ac.cy).}}
\maketitle

\begin{abstract}
The emerging reconfigurable antenna (RA) array technology promises capacity enhancement through dynamic antenna positioning. Traditional approaches enforce half-wavelength or greater spacing among RA elements to avoid mutual coupling, limiting the solution space. Additionally, achieving sufficient spatial channel sampling requires numerous discrete RA positions (ports), while high-frequency scenarios with hybrid processing demand many physical RAs to maintain array gains. This leads to exponential growth in the solution space. In this work, we propose two techniques to address the former challenge: (1) surrounding a limited number of active RAs with passive ones terminated to tunable analog loads to \textit{exploit} mutual coupling and increase array gain, and (2) employing tunable loads on each RA in an all-active design to \textit{eliminate} mutual coupling in the analog domain. Both methods enable arbitrary RA spacing, unlocking the full solution space. Regarding the latter challenge, we develop greedy and meta-heuristic port selection algorithms, alongside low-complexity heuristic variants, that efficiently handle over $10^{20}$ array configurations. Furthermore, we optimize the loading values to maximize the sum-rate in a multiple-input single-output broadcast channel under transmission power constraints, assuming a heuristic linear precoder. In addition, we analyze performance degradation from quantized loads and propose corresponding robust designs. Numerical simulations reveal 20-56\% sum-rate gains over benchmarks and around 60\% performance recovery under quantization errors.
\end{abstract}

\begin{IEEEkeywords}
MIMO, parasitic antennas, reconfigurable antennas, quantized loads, mutual coupling, optimization.
\end{IEEEkeywords}

\section{Introduction}\label{sec:1}
The advent of 5th Generation (5G) networks has been driven by envisioned applications such as augmented/virtual reality and smart grids requiring enhanced capacity and user throughput, massive connectivity, and improved reliability~\cite{5GUseCases,5GKPIs}. 5G leverages massive multiple-input multiple-output (M-MIMO) technology, which employs an excess of antennas, and underutilized millimeter-wave (mmWave) spectrum to meet these requirements, offering substantial spatial multiplexing and beamforming gains while expanding spectral bandwidth~\cite{SIG-093,mmWavemMIMOSurv}. Nevertheless, fully-digital transceivers are impractical at mmWave frequencies due to their high hardware cost, power consumption, and complexity, as each antenna requires a dedicated radio frequency (RF) chain.

Hybrid analog-digital transceivers address these challenges by using fewer RF units, which are connected to the antenna array via an analog beamforming network consisting of phase shifters, and, consequently, splitting pre-/post-processing among the digital and analog domains. The hybrid fully-connected (HFC) structure approaches fully-digital spectral efficiency but suffers from notable power consumption overhead, while the hybrid sub-connected (HSC) design is more power-efficient but incurs non-negligible performance losses~\cite{HybBFSurv}. These trade-offs motivate alternative architectures.

Parasitic antenna arrays exploit the strong mutual coupling between a few active (RF-fed) antennas and nearby passive elements terminated with tunable analog loads (e.g., varactors), enabling reduced hardware cost and power consumption and better spectral/energy efficiency balance than the above hybrid designs~\cite{ESPARbook,Heath2025}. Load impedance control determines the induced currents on passive antennas via coupling, enabling parasitic radiation. However, such multiple-active multiple-passive (MAMP) arrays face challenges, including impedance-dependent dynamic matching and complex current control, which have limited their adoption. Hence, prior works focused on single-fed arrays, also known as electronically steerable parasitic array radiators (ESPARs), for compact, low-power devices like phones and sensors~\cite{BMIMOESPAR,AdaptESPARBMIMO,ESPARArbPrec,SingleRFPA,ESPARQuantLoads}, leaving the study of MAMP arrays for high-capacity use cases underexplored.

Controlling mutual coupling via tunable loads has also been proposed for fully-digital arrays, by connecting the antennas to their RF chains through analog loads, either to suppress coupling~\cite{TunLoads} or exploit it under symbol-level precoding~\cite{TunLoads2}.

As the industry transitions toward 6G, driven by services with extreme performance demands (e.g., holographic communication, ultra-massive IoT~\cite{6G,WangRoad}) and sustainability concerns (e.g., energy crisis, climate change~\cite{Green6G}), MIMO must evolve. Extra-large MIMO (XL-MIMO) expands M-MIMO scale by an order of magnitude~\cite{XLMIMO}, while holographic MIMO employs quasi-continuous electromagnetic surfaces with sub-wavelength meta-atoms~\cite{HMIMOSurv}. Both promise unprecedented spatial degrees-of-freedom (DoF) and near-field focusing at mmWave/THz bands, but pose major dimensionality-related signal processing and channel estimation challenges.

Reconfigurable antenna (RA) arrays are a next-generation MIMO paradigm, wherein antennas are dynamically repositioned within a confined region to enhance spatial diversity and unlock new spatial DoF using fewer physical antennas and RF chains~\cite{FLPMIMO}. Their ability to adapt to channel variations, creating favorable propagation conditions, or alter the radiation pattern by modifying their geometry is unmatched by fixed-position antenna (FPA) arrays and antenna selection (AS)~\cite{XL-MIMO-AS}.

RA arrays include movable antenna (MA) and fluid antenna (FA) designs. MA arrays use motor-driven metallic antennas~\cite{MAMIMO,MAS,MAS2}, while FAs employ liquid-metal antennas (e.g., EGaIn, Galinstan) whose position is controlled via electric fields or micro-pumps~\cite{FAS,FAS2,FAS3}. MA designs facilitate 2D/3D configurations but require higher power and cost than FAs due to the mechanical drivers. FA arrays also offer faster reconfiguration and support shape/volume tuning. Both architectures theoretically enable antenna positioning over a continuum for fine spatial channel sampling, yet practical limitations (e.g., finite resolution) promote implementations with predetermined discrete sub-wavelength-spaced possible antenna positions (ports), particularly for FAs. Thus, MA/FA arrays are typically modeled as continuous/discrete, respectively~\cite{MAMIMO,FAS2}. Nonetheless, both technologies are mathematically equivalent under identical RA positioning.

\subsection{Related Works}\label{subsec:1.1}
Research on FA/MA systems has gained attention lately. The authors in~\cite{FAS,FAMRC} show that single-FA receivers with a sufficient number of discrete ports outperform FPA-based maximum ratio combining in terms of outage probability and ergodic capacity, respectively, whereas the work in~\cite{MAS} demonstrates the considerable gains of single-MA systems over conventional FPA systems with/without AS.

In uplink communication scenarios, FA multiple access improves network outage probability through port selection at single-FA user terminals for interference mitigation purposes~\cite{FAMA}, while~\cite{ULPGD,MAS2} minimize transmit power via antenna position optimization at FA/MA-equipped base stations (BS) adopting the zero-forcing (ZF) or minimum mean square error receiver architecture. In~\cite{FACFMIMO}, uplink transmit powers and FA positions at single-FA access points are jointly optimized to maximize the minimum uplink spectral efficiency in cell-free massive MIMO networks. The work in~\cite{HybFPAMA} proposes a novel transceiver architecture combining FPAs with 3D position-/rotation-adjustable MA surfaces and investigates uplink capacity maximization via rotation angle optimization.

The joint optimization of transmit/receive FA positions to maximize the achievable rate in single-user MIMO systems is explored in~\cite{MIMOFA}, while the authors in~\cite{MAMIMO} optimize along the transmit covariance matrix, assuming MAs-equipped nodes. Extensions utilizing statistical, instead of instantaneous, channel state information have been developed for both FAs-~\cite{FAMIMOSCSI} and MAs-based~\cite{MAMIMOSCSI} configurations. Moreover, the works in ~\cite{FADLTSP,MATSP} study transmit sum-power minimization in multi-user multiple-input single-output (MISO) downlink systems considering either a FPAs- or MAs-equipped BS and single-FA or single-FPA user terminals, respectively. The authors jointly optimize the precoder and the discrete port selection of the receive FAs or transmit MAs, respectively.~\cite{FlexPrecMA} introduces a flexible MAs-enhanced precoding method that utilizes regularized least squares-orthogonal matching pursuit.~\cite{MAWSRMax,FAMISOBCSR} tackle weighted or non-weighted sum-rate maximization via joint precoding and transmit MA/FA positions optimization, respectively, whereas~\cite{zhang2024movable} presents a respective hybrid precoding design for MA-equipped BSs.

More recently, the authors in~\cite{RISMA} focus on the synergy between the MA and reconfigurable intelligent surface (RIS) technologies by handling sum-rate maximization through the joint optimization of precoding, RIS beamforming, and MA positions, while~\cite{FAHybRIS} considers an FA-equipped BS assisted by a hybrid sub-connected active/passive RIS and optimizes RIS elements' scheduling to maximize the energy efficiency.

\subsection{Motivation, Goals, and Contributions}\label{subsec:1.2}
Despite their advantages, FA/MA systems face also several challenges: (1) antenna spacing is restricted to half-wavelength or greater to avoid the occurrence of mutual coupling, thus limiting the solution space; (2) antenna positions optimization approaches present high computational complexity that discourages practical application, since continuous implementations typically involve some gradient descent variant while discrete realizations give rise to combinatorial problems with large search spaces, due to the large number of ports required for sufficient performance gains; (3) the requirement to estimate the user channels with all potential antenna positions introduces significant channel estimation overhead; (4) optimization and processing complexity increase in high-frequency (mmWave/THz) scenarios, wherein a large number of physical antennas is necessary to achieve sufficient beamforming gains that adequately compensate for the severe path loss; (5) in such setups, conventional hybrid processing structures, which yield suboptimal spectral-/energy-efficiency trade-off, are utilized to reduce the number of RF chains.

This work considers a discrete linear RA array, i.e., an array of metallic or liquid-metal antennas with positions adjusted in discrete steps (ports), which is more practical to implement. We study a MISO broadcast channel, aiming to maximize the sum-rate under a transmit sum-power constraint using a heuristic linear precoder. To address the outlined challenges, a synergy of techniques is introduced. Notably, we propose two transceiver designs: (1) a MAMP RA array for mmWave M-MIMO that \textit{exploits} mutual coupling to enhance array gain and improve spectral and energy efficiency over current hybrid designs, and (2) an all-active RA array that connects all antennas to RF chains and employs tunable analog loads to \textit{eliminate} coupling in the analog domain. We note that both architectures allow arbitrary RA spacing, thereby unlocking the full design space. Note that varactors and other analog loads support mmWave operation, as demonstrated in practice~\cite{Var}.

Given these transceivers, sum-rate maximization entails optimization of RA positions and load values. We develop a greedy RA port selection (RAPS) algorithm and adapt binary particle swarm optimization (BPSO) and tabu search (TS) meta-heuristics to efficiently handle over $10^{20}$ possible array configurations. This is crucial for MAMP arrays, which use more antennas to achieve greater array gain. Also, heuristic RAPS algorithms for the MAMP design are proposed to further reduce complexity and channel estimation overhead, and analytical solutions for the optimal loading values are derived. The performance loss due to quantized loads is analyzed, and corresponding robust designs are developed.

We should mention the following: (1) Efficient channel estimation for discrete RA (e.g., FA) arrays~\cite{FAsChEst} and impedance matching in single-/multi-RF parasitic arrays~\cite{ESPARArbPrec} have been addressed in the literature and lie outside this paper’s scope. (2) As the first work exploring the integration of RA and parasitic array technologies, our goal is to unveil trends, provide insights, and derive performance bounds. Thus, we assume perfect channel knowledge at the BS to maintain mathematical tractability, leaving the imperfect channel state information scenario for future investigation.

The key contributions of this study are summarized below:
\begin{itemize}
\item Two novel transceiver designs employing the integration of RA and parasitic antenna technologies for the first time, with opposite operational principles (exploiting versus eliminating mutual coupling), are introduced to unlock arbitrary antenna spacing, unlike existing frameworks.
\item A greedy RAPS algorithm is proposed, and two meta-heuristics (BPSO, TS) are adapted to exponentially reduce complexity from exhaustive search $\mathcal{O}(N^M)$ while efficiently handling $>10^{20}$ array configurations.
\item Heuristic variants of these algorithms are designed to further reduce complexity in the MAMP RA array scenario.
\item A practical decoupled RAPS/loads optimization approach is followed, based on the problem structure. The loading values are optimized, for the first time in the case of a MAMP array, to the best of our knowledge, assuming the use of some heuristic linear precoder.
\item The performance degradation caused by quantized loads is analyzed, and robust designs are developed. 
\end{itemize}
Comprehensive numerical simulations across various test scenarios unveil 20-56\% sum-rate gains over benchmarks and 55-60\% performance recovery under quantization errors.

\subsection{Structure and Mathematical Notation}\label{subsec:1.3}
\begin{figure}[t!]
\centering
\includegraphics[width=\columnwidth]{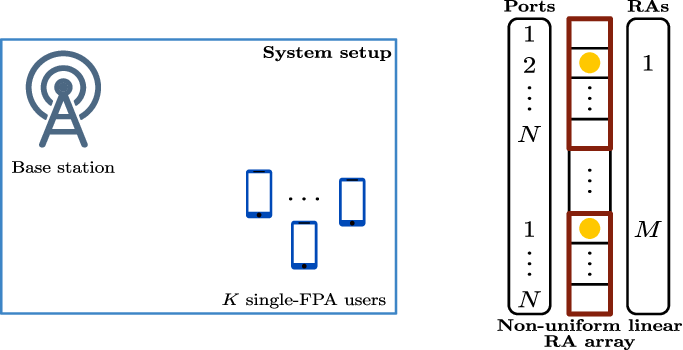}
\caption{System setup and proposed RA array structure.}
\label{fig:1}
\end{figure}
The rest of the paper is organized as follows: The system and channel models are introduced in Section II. In Section III, the optimization problem for RAPS is formulated, and respective designs are developed. Section IV deals with load impedance optimization for the MAMP and all-active, tunable loads-assisted RA arrays. Section V is devoted to performance analysis under quantized loads and presents corresponding robust designs. In Section VI, the performance of the proposed schemes is comparatively evaluated against benchmarks via numerical simulations, while Section VII concludes the paper.

\textbf{Notation:} $\mathbf{a}$ is column vector, $a_i$ denotes its $i$-th element, $\left\|\mathbf{a}\right\|$ and $\left\|\mathbf{a}\right\|_0$ respectively represent its $\ell_2$ and $\ell_0$ norm. $\mathbf{A}$ is a matrix, $A_{i,j}$ denotes its $(i,j)$-th entry, $\mathbf{A}^{-1}$, $\mathbf{A}^T$, $\mathbf{A}^*$, $\mathbf{A}^{\dagger}$, $\mathbf{A}^{+}$ $\left\|\mathbf{A}\right\|_F$, and $\operatorname{Tr}\left(\mathbf{A}\right)$ respectively represent its inverse, transpose, complex conjugate, conjugate transpose, Moore-Penrose pseudo-inverse, Frobenius norm, and trace. $\left[\mathbf{a}_1;\dots;\mathbf{a}_M\right]$ denotes the vertical stacking of vectors. $\operatorname{diag}\left(\mathbf{a}\right)$ represents the conversion of $\mathbf{a}$ to a diagonal matrix. $\operatorname{blkdiag}\left(\mathbf{a}_1,\dots,\mathbf{a}_M\right)$ stands for a block diagonal matrix. $\mathbf{I}_N$ refers to the $N\times N$ identity matrix. $\mathbb{E}\left[\cdot\right]$, $a \bmod b$, and $\lfloor \cdot \rfloor$ respectively correspond to the expectation, modulo, and floor operations. $\mathcal{CN}\left(\cdot,\cdot\right)$ and $\mathcal{U}\left(a,b\right)$ respectively denote the complex Gaussian and uniform distributions. $\left|a\right|$ is the magnitude of $a\in\mathbb{C}$. $\left|\mathcal{A}\right|$ denotes the cardinality of set $\mathcal{A}$. Also, $\mathbb{B}\triangleq\{0,1\}$ and $\mathbb{R}_+\triangleq[0,+\infty)$.

\section{System and Channel Models}\label{sec:2}
In the considered system, a BS equipped with an $M$-RAs linear array serves $K$ single-FPA user terminals. As shown in Fig.~\ref{fig:1}, there are $N$ possible predefined discrete positions (ports) per RA, resulting in $L = NM$ ports in total, $L \gg M$, and $N^M$ array configurations ($M$-tuples of activated ports).

Let $\mathcal{P}_{m}\triangleq\left\{x_{m,1},\dots,x_{m,N}\right\}$ hold $m$-th RA's port positions, where $x_{m,n}\in\mathbb{R}_{+}$ is the position of the $n$-th port, $m\in\mathcal{M}\triangleq\left\{1,\dots,M\right\}$, $n\in\mathcal{N}\triangleq\left\{1,\dots,N\right\}$, and $\mathcal{P}=\left\{x_{1,1},\dots,x_{1,N},\dots,x_{M,1},\dots,x_{M,N}\right\}\triangleq\left\{x_1,\dots,x_L\right\}$ hold all $L$ ports' positions, where $x_l\in\mathbb{R}_{+}$ is the position of the $l$-th port, $l\in\mathcal{L}\triangleq\{1,\dots,L\}$. By setting $x_{1,1} = x_1 = 0$, $x_{m,n} = \left[(m-1)N+n-1\right]d=(l-1)d = x_l$, where $d=\bar{d}/\lambda$ is the port spacing $\bar{d}$ normalized to the wavelength $\lambda=c/f_c$, $f_c$ is the carrier frequency and $c=3\times 10^8$ m/s is the speed of light. Furthermore, we define the $m$-th RA's position and array configuration as $t_m\in\mathcal{P}_m$ and $\mathbf{t}=\left[t_1,\dots,t_M\right]\in\mathbb{R}_{+}^M$.

\subsection{MAMP RA Array}\label{subsec:2.1}
\subsubsection{System Setup and Channel Model}\label{subsubsec:2.1.1}
\begin{figure}[t!]
\centering
\includegraphics[width=\columnwidth]{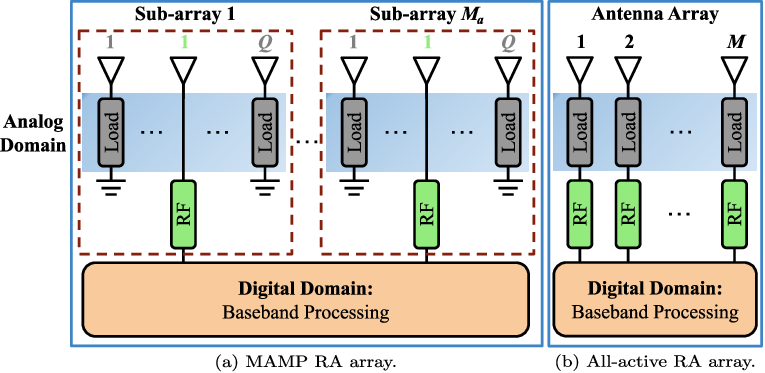}
\caption{Proposed RAs-equipped arrays with tunable loads.}
\label{fig:2}
\end{figure}

In the transceiver with the MAMP array, the RAs are divided into $M_a$ active and $M_p = M-M_a=Q M_a$ passive ones terminated to tunable analog loads, $Q\in\mathbb{N}$. Contrary to the all-active variant, the array is partitioned into $M_a$ sub-arrays with $M_p/M_a = Q$ passive and 1 active RA each, as illustrated in Fig.~\ref{fig:2}(a).

The channel from all $N$ ports of the $m$-th RA to the $k$-th user is $\widehat{\mathbf{h}}_{m,k}^{\dagger}\left(\mathbf{p}_m\right)=\left[h_{m,1,k}^*\left(x_{m,1}\right),\dots,h_{m,N,k}^*\left(x_{m,N}\right)\right]\in\mathbb{C}^N$, where $\mathbf{p}_m \triangleq \left[x_{m,1},\dots,x_{m,N}\right]^T\in\mathbb{R}_{+}^N$ and $h_{m,n,k}^*\left(x_{m,n}\right)\in\mathbb{C}$ corresponds to the channel from the $n$-th port, $k\in\mathcal{K}\triangleq\{1,\dots,K\}$. Hence, the channel from all $L$ ports to the $k$-th user is denoted by $\widehat{\mathbf{h}}_k^{\dagger}\left(\mathbf{p}\right)=\left[\widehat{\mathbf{h}}_{1,k}^{\dagger}\left(\mathbf{p}_1\right),\dots,\widehat{\mathbf{h}}_{M,k}^{\dagger}\left(\mathbf{p}_M\right)\right]\in\mathbb{C}^{L}$, where $\mathbf{p}=\left[\mathbf{p}_1;\dots;\mathbf{p}_M\right]\triangleq\left[x_1,\dots,x_L\right]^T\in\mathbb{R}_{+}^L$. According to the geometric Saleh-Valenzuela channel model~\cite{SSP}
\vspace{-1mm}
\begin{equation}\label{eq:1}
\widehat{\mathbf{h}}_k\left(\mathbf{p}\right) = \sqrt{\frac{L}{N_c N_p}} \sum_{c\in\mathcal{N}_c} \sum_{p\in\mathcal{N}_p} \beta_{c, p, k} \ \widehat{\mathbf{a}}_k\left(\hat{\theta}_{c, p, k}\right),
\vspace{-1mm}
\end{equation} 
where $N_c$ and $N_p$ is the number of scattering clusters and paths per cluster, respectively ($N_c N_p$ paths in total), $\beta_{c, p, k}$ denotes the complex gain of the $p$-th path in the $c$-th cluster, $p\in\mathcal{N}_p\triangleq\left\{1,\dots,N_p\right\}$, $c\in\mathcal{N}_c\triangleq\left\{1,\dots,N_c\right\}$, and $\widehat{\mathbf{a}}_k\left(\hat{\theta}_{c, p, k}\right)\in\mathbb{C}^L$ represents the steering vector of the BS, which is given by  
\begin{equation}\label{eq:2}
\widehat{\mathbf{a}}_k\left(\hat{\theta}_{c,p,k}\right)=\frac{1}{\sqrt{L}}\left[e^{-j 2 \pi \hat{\theta}_{c,p,k}^{(1)}}, \dots, e^{-j 2 \pi \hat{\theta}_{c,p,k}^{(L)}}\right]^T.
\vspace{-1mm}
\end{equation}
In Eq.~\eqref{eq:2}, $\hat{\theta}_{c,p,k}^{(l)}=x_l \cos \left(\hat{\phi}_{c,p,k}\right) = (l-1)d\cos \left(\hat{\phi}_{c,p,k}\right)$ are the spatial frequencies and $\hat{\phi}_{c,p,k} \in[-\pi / 2, \pi / 2]$ are the angles-of-departure (AoD) measured relative to the azimuth.

\subsubsection{RAs Position Selection and Heuristic Linear Precoders}\label{subsubsec:2.1.3}
RA port selection (RAPS) is performed based on perfect knowledge of $\widehat{\mathbf{h}}_k^{\dagger}\left(\mathbf{p}\right)$ at the BS. Let $\mathbf{b}_{m}\triangleq\left[b_{m,1},\dots,b_{m,N}\right]^T\in\mathbb{B}^N$ be a binary selection vector for the $m$-th RA, $b_{m,n}\in\{0,1\}$. If this RA occupies its $n$-th port, then $b_{m,n}=1$; otherwise, $b_{m,n}=0$. Since an RA can occupy only one port, $\sum_{n=1}^{N}b_{m,n}=1$. By defining the binary selection matrix (BSM) $\bm{\Sigma}\triangleq\operatorname{blkdiag}\left(\mathbf{b}_1,\dots,\mathbf{b}_M\right)\in\mathbb{B}^{L\times M}$,
\begin{equation}\label{eq:3}
\mathbf{h}_k\left(\mathbf{t}\right) = \bm{\Sigma}^T\widehat{\mathbf{h}}_k\left(\mathbf{p}\right),
\end{equation}  
where $\mathbf{h}_k^{\dagger}\left(\mathbf{t}\right)\in\mathbb{C}^M$ represents the channel from all $M$ RAs at the selected positions (ports) to the $k$-th user and $\mathbf{t}$ is the array configuration vector. $\mathbf{h}_k\left(\mathbf{t}\right)$ is described by Eq.~\eqref{eq:1} by replacing $L$ with $M$ and using steering vectors $\mathbf{a}_k\left(\theta_{c,p,k}\right)$ with spatial frequencies $\theta_{c,p,k}^{(m)}=t_m\cos\left(\phi_{c,p,k}\right)=x_{m,n}\cos\left(\phi_{c,p,k}\right)= \left[(m-1)N+n-1\right]d\cos\left(\phi_{c,p,k}\right)$, $\phi_{c,p,k}\in\left[-\pi/2,\pi/2\right]$. The channel from all $M$ RAs to all $K$ users is constructed as $\mathbf{H}^{\dagger}\left(\mathbf{t}\right) = \left[\mathbf{h}_1\left(\mathbf{t}\right),\dots,\mathbf{h}_K\left(\mathbf{t}\right)\right]^{\dagger}\in\mathbb{C}^{K\times M}$.

Let's consider for a moment a BS with $M$ active RAs (without analog loads). We assume the application of linear precoding schemes. The transmitted signal, $\mathbf{x}\left(\mathbf{H}\right)\in\mathbb{C}^M$, is
\begin{equation}\label{eq:9}
\mathbf{x}\left(\mathbf{H}\right) = \sum_{k\in\mathcal{K}}\mathbf{w}_k\left(\mathbf{h}_k\right) s_k = \mathbf{W}\left(\mathbf{H}\right)\mathbf{s},
\end{equation} 
where $\mathbf{w}_k\left(\mathbf{h}_k\right)\in\mathbb{C}^M$ denotes the precoding vector for the $k$-th user and $\mathbf{W}\left(\mathbf{H}\right)=\left[\mathbf{w}_1\left(\mathbf{H}\right),\dots,\mathbf{w}_{K}\left(\mathbf{H}\right)\right]\in\mathbb{C}^{M\times K}$ refers to the precoding matrix. Moreover, $s_k\sim\mathcal{CN}(0,1)$ represents the independent and identically distributed (i.i.d.) symbol transmitted to user $k$ and $\mathbf{s}=\left[s_1,\dots,s_K\right]^T\in\mathbb{C}^K$ stands for the i.i.d. transmitted symbols vector, with $\mathbb{E}\left[\mathbf{s}\mathbf{s}^{\dagger}\right]=\mathbf{I}_K$. The transmit sum power $P_t = \!\mathbb{E}\left[\left\|\mathbf{x}\left(\mathbf{H}\right)\right\|^2\right]$, is constrained as
\begin{equation}\label{eq:10}
P_t = \sum_{k\in\mathcal{K}}\left\|\mathbf{w}_k\left(\mathbf{H}\right)\right\|^2 = \operatorname{Tr}\left(\mathbf{W}^{\dagger}\left(\mathbf{H}\right)\mathbf{W}\left(\mathbf{H}\right)\right)\leq P_{\max},
\end{equation}
where $P_{\max}\geq 0$ is the transmit power budget.

We consider the heuristic matched filter (MF), ZF, and Wiener filter (WF) linear precoders, which are defined as
\begin{align}\label{eq:15}
\mathbf{W} &= \alpha\mathbf{F} = \alpha\left[\mathbf{f}_1,\dots,\mathbf{f}_K\right], \ \alpha = \sqrt{\frac{P_{\max}}{\operatorname{Tr}\left(\mathbf{F}\mathbf{F}^{\dagger}\right)}},
\end{align}
where $\mathbf{F}_{\text{MF}} = \mathbf{H}$, $\mathbf{F}_{\text{ZF}} = \mathbf{H}^{+} = \mathbf{H}\left[\mathbf{H}^{\dagger}\mathbf{H}\right]^{-1}$, and $\mathbf{F}_{\text{WF}} = \left[\mathbf{H}\mathbf{H}^{\dagger}+\zeta\mathbf{I}_K\right]^{-1}\mathbf{H}, \ \zeta = \frac{\sigma^2 K}{P_{\max}}$. $\mathbf{F}\in\mathbb{C}^{M\times K}$ is the non-normalized precoding matrix, $\mathbf{f}_k\in\mathbb{C}^{M}$ refers to the respective non-normalized precoding vector for the $k$-th user, $\alpha$ denotes a normalization factor which ensures that the transmit sum-power constraint in Eq.~\eqref{eq:10} is satisfied, and we dropped functional dependencies for brevity. We note that we use representative linear precoders to evaluate candidate RA positions, which is essential for port selection search over a large discrete space. Indeed, once the effective channel is formed for a given configuration, any digital preoding method can be applied. Incorporating nonlinear precoding schemes is an interesting direction and left for future consideration.

\subsubsection{Signal Model and Mutual Coupling}\label{subsubsec:2.1.4}
For convenience and without loss of generality, we define the index sets of active and passive RAs as $\mathcal{M}_a\triangleq\left\{1,\dots,M_a\right\}$ and $\mathcal{M}_p\triangleq\left\{M_a + 1,\dots,M\right\}$, respectively. As illustrated in the equivalent circuit of a MAMP array in Fig.~\ref{fig:3}, each active RA is fed by an independent source with complex voltage $v_{m}$ and output impedance $z_0$, $m\in\mathcal{M}_a$, while each passive RA is terminated to a load with impedance $z_m$, $m\in\mathcal{M}_p$. According to generalized Ohm's law, the electric currents on the RAs, $\mathbf{i}\left(\mathbf{t},\mathbf{z}_L\right)=\left[i_1,\dots,i_M\right]^T\in\mathbb{C}^M$, are related with the source voltages, $\mathbf{v}=\left[v_1,\dots,v_{M_a},0,\dots,0\right]\in\mathbb{C}^{M}$, as follows~\cite{ESPARQuantLoads}
\begin{equation}\label{eq:4}
\mathbf{i}\left(\mathbf{t},\mathbf{z}_L\right) = \mathbf{Z}_T\left(\mathbf{t},\mathbf{z}_L\right)\mathbf{v},
\end{equation}
where $\mathbf{Z}_T\left(\mathbf{t},\mathbf{z}_L\right)\triangleq\left(\mathbf{Z}\left(\mathbf{t}\right)+\mathbf{X}\right)^{-1}\in\mathbb{C}^{M\times M}$ is the effective coupling matrix relating voltages to currents, $\mathbf{X}\triangleq\operatorname{diag}\left(\mathbf{z}_L\right)\in\mathbb{C}^{M\times M}$, $\mathbf{z}_L = \left[z_0,\dots,z_0,z_{M_a+1},\dots,z_M\right]^T\in\mathbb{C}^M$ holds the source and analog load impedances, and $\mathbf{Z}\left(\mathbf{t}\right)\in\mathbb{C}^{M\!\times\! M}$ stands for the mutual impedance matrix, which is computed via the induced EMF method~\cite{balanis} (see Appendix A). $Z_{m,m}=z_A$ holds the self-impedance of the respective RA and $Z_{m,m^{\prime}}$, $m^{\prime}\in\mathcal{M}\setminus\{m\}$, holds the mutual impedance between the $m$-th and $m^{\prime}$-th RA. The former depends on antenna length and carrier frequency, while mutual impedances depend on the normalized RA distances. $\mathbf{Z}\left(\mathbf{t}\right)$ is symmetric, since $Z_{m,m^{\prime}} = Z_{m^{\prime},m}$ due to reciprocity. Its overall structure depends on the array geometry. For a uniform linear array, wherein antenna spacing is constant, the result is a Toeplitz matrix. This is not the case in the considered non-uniform linear array, where the distance between two RAs depends on port selection.

To account for mutual coupling, we define $\widebar{\mathbf{H}}^{\dagger}\left(\mathbf{t},\mathbf{z}_L\right)=\left[\widebar{\mathbf{h}}_1\left(\mathbf{t},\mathbf{z}_L\right),\dots,\widebar{\mathbf{h}}_K\left(\mathbf{t},\mathbf{z}_L\right)\right]^{\dagger}\in\mathbb{C}^{K\times M}$ as~\cite{BMIMOESPAR}
\begin{subequations}\label{eq:7}
\begin{alignat}{2}
\widebar{\mathbf{H}}^{\dagger}\left(\mathbf{t},\mathbf{z}_L\right)&=\mathbf{H}^{\dagger}\left(\mathbf{t}\right)\mathbf{C}\left(\mathbf{t},\mathbf{z}_L\right), \label{eq:7a} \\[2mm]
\mathbf{C}\left(\mathbf{t},\mathbf{z}_L\right) &= \left(z_A\mathbf{I}_M+\mathbf{X}\right)\mathbf{Z}_T\left(\mathbf{t},\mathbf{z}_L\right), \label{eq:7b}
\end{alignat}
\end{subequations}
where $\mathbf{C}\left(\mathbf{t},\mathbf{z}_L\right)\in\mathbb{C}^{M\times M}$ is the mutual coupling matrix, transforming the impedance-based current relationships into channel-based signal relationships and capturing coupling-induced power degradation, which is minimized via load impedance optimization. The precoder is subsequently adapted to the channel matrix $\widebar{\mathbf{H}}^{\dagger}\left(\mathbf{t},\mathbf{z}_L\right)$ according to Eq.~\eqref{eq:15}.

\subsubsection{System Model}\label{subsubsec:2.1.5}
Let's return to our all-active RA array (without analog loads) example. In this case, $\widebar{\mathbf{H}}$ depends solely on $\mathbf{t}$. The received signal at the $k$-th user is expressed as
$y_k = \widebar{\mathbf{H}}^{\dagger}\left(\mathbf{t}\right)\mathbf{W}\mathbf{s} + n_k = \widebar{\mathbf{h}}_k^{\dagger}\left(\mathbf{t}\right)\mathbf{w}_k s_k + \widebar{\mathbf{h}}_k^{\dagger}\left(\mathbf{t}\right)\sum_{i\in\mathcal{K}\setminus{k}}\mathbf{w}_i s_i + n_k$,
where $n_k\sim\mathcal{CN}(0,\sigma^2)$ represents the additive white Gaussian noise (AWGN) at user $k$.

\begin{figure}[t!]
\centering
\includegraphics[width=\columnwidth]{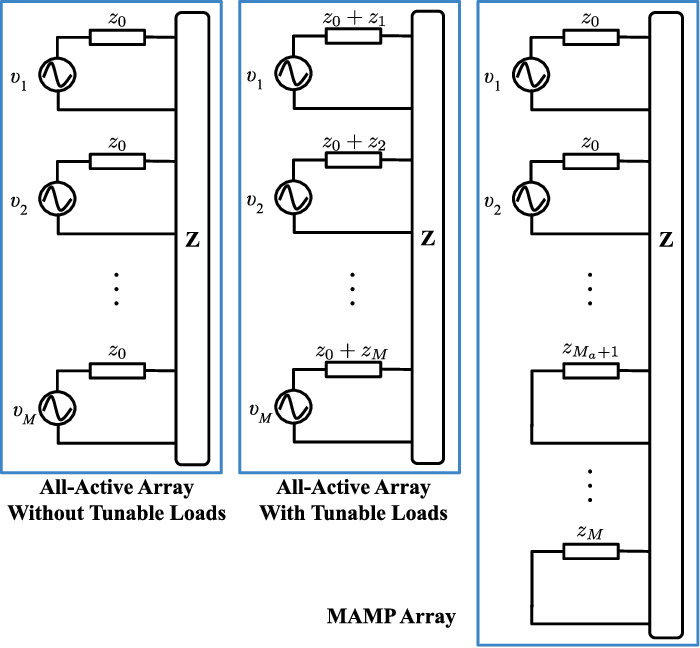}
\caption{Equivalent circuit representation of the all-active and MAMP arrays.}
\label{fig:3}
\end{figure}

The equivalent circuit of an all-active array (without analog loads) is shown in Fig.~\ref{fig:3}. Generalized Ohm's law applies with $\mathbf{v}=\mathbf{v}_0=\left[v_1,\dots,v_M\right]^T\in\mathbb{C}^M$ and $\mathbf{X}=\mathbf{X}_0=z_0\mathbf{I}_M$. With antenna spacing of $\lambda/2$ or greater to prevent mutual coupling, $z_A = z_0^*$ to ensure maximum power transfer, yielding $\mathbf{Z}\left(\mathbf{t}\right)=z_0^*\mathbf{I}_M$ and $\mathbf{i}\left(\mathbf{t}\right)=\mathbf{i}_0\left(\mathbf{t}\right)=\left(z_0+z_0^*\right)\mathbf{v}_0$~\cite{ESPARQuantLoads}.

The system model can be written as $\mathbf{y} = \widebar{\mathbf{H}}^{\dagger}\left(\mathbf{t}\right)\mathbf{W}\mathbf{s} + \mathbf{n}$, where $\mathbf{y}=\left[y_1,\dots,y_K\right]^T\in\mathbb{C}^K$ and $\mathbf{n}\sim\mathcal{CN}\left(\mathbf{0},\sigma_k^2\mathbf{I}_K\right)$ are the received signals and AWGN vectors, respectively. In fully-active arrays, the precoded signal $\mathbf{x}=\mathbf{W}\mathbf{s}$ corresponds to $M$ source voltages $\mathbf{v}_0$, translated to RA currents $\mathbf{i}_0$ that determine each RA's response and, therefore, the array's radiation pattern. Thus, a scaled channel matrix relates channel input (antenna currents $\mathbf{i}_0$) to channel output (open-circuit voltages at user terminals $\mathbf{y}$): $\mathbf{y} = \widebar{\mathbf{H}}^{\dagger}\left(\mathbf{t}\right)\mathbf{v}_0 + \mathbf{n} = \frac{1}{z_0+z_0^*}\widebar{\mathbf{H}}^{\dagger}\left(\mathbf{t}\right)\mathbf{i}_0 + \mathbf{n}$~\cite{ESPARQuantLoads}.

For MAMP arrays, the precoded signal $\mathbf{x}_a=\mathbf{W}_a\mathbf{s}\in\mathbb{C}^{M_a}$, where $\mathbf{W}_a\in\mathbb{C}^{M_a\times K}$, generates $M_a$ source voltages that create currents in $M_a$ active RAs and induce currents in $M_p = M-M_a$ passive RAs. The channel matrix relates user terminal voltages to antenna currents as $\mathbf{y} = \widebar{\mathbf{H}}^{\dagger}\left(\mathbf{t},\mathbf{z}_L\right)\mathbf{i} + \mathbf{n}$. For given precoding and modulation schemes, source voltages and load impedances can be adjusted such that $\mathbf{i}=\left(z_0+z_0^*\right)^{-1}\mathbf{i}_0=\mathbf{v}_0=\mathbf{W}\mathbf{s}$, to resemble an all-active RA array without loads. 

\subsection{All-Active RA Array With Tunable Loads}\label{subsec:2.2}
In this architecture, we employ tunable loads on each RA in an all-active design, as depicted in Fig.~\ref{fig:2}(b). More specifically, each RA connects to its RF chain via a tunable load with impedance $z_m$, as shown in Fig.~\ref{fig:3}, resulting in $\mathbf{z}_L = \left[z_0+z_1,\dots,z_0+z_M\right]^T\in\mathbb{C}^M$.

The MAMP RA array mainly targets high-band M-MIMO communication scenarios ($> 24$ GHz). The all-active RA array with tunable loads, in turn, mainly targets mid-band (1 GHz -- 6 GHz) and centimeter-wave (cmWave) spectrum (7 GHz -- 15 GHz) applications. Hence, the channel between all $L$ ports and user $k$ is modeled as~\cite{TunLoads}
\begin{equation}\label{eq:AA1}
\widetilde{\mathbf{h}}_k^{\dagger}\left(\mathbf{p}\right) = \mathbf{g}_k^{\dagger}\widetilde{\mathbf{A}}_k,
\end{equation} 
where $\mathbf{g}_k^{\dagger}\in\mathbb{C}^{N_r}$ is the Rayleigh fading component with $\mathbf{g}_k^{\dagger}(r)\sim\mathcal{CN}(0,1)$, $r\in\mathcal{N}_r \triangleq\left\{1,\dots,N_r\right\}$, $N_r$ denotes the number of directions-of-departure, $\widetilde{\mathbf{A}}_k = \left[\widetilde{\mathbf{a}}^T\left(\tilde{\varphi}_{k,1}\right),\dots,\widetilde{\mathbf{a}}^T\left(\tilde{\varphi}_{k,N_r}\right)\right]^T\in\mathbb{C}^{N_r\times L}$ contains the $N_r$ steering vectors at the BS side that model the spatial correlation, $\tilde{\varphi}_{k, r}$ are the AoD, and $\widetilde{\mathbf{a}}\left(\hat{\varphi}_{k, r}\right)\in\mathbb{C}^L$ are given by
\begin{equation}\label{eq:AA2}
\widetilde{\mathbf{a}}\left(\tilde{\varphi}_{k, r}\right)=\left[1, e^{j 2 \pi d \sin \tilde{\varphi}_{k, r}}, \ldots, e^{j 2 \pi\left(L-1\right) d \sin \tilde{\varphi}_{k, r}}\right].
\end{equation} 
After RAPS, we obtain $\mathbf{h}_k^{\dagger}(\mathbf{t})\in\mathbb{C}^M$ according to Eq.~\eqref{eq:3}. To account for mutual coupling, we define $\widebar{\mathbf{H}}^{\dagger}\left(\mathbf{t},\mathbf{z}_L\right)\in\mathbb{C}^{K\times M}$ according to Eq.~\eqref{eq:7}. The received signal vector is given by
\begin{align}\label{eq:AA4}
\mathbf{y} &= \widebar{\mathbf{H}}^{\dagger}\left(\mathbf{t},\mathbf{z}_L\right)\mathbf{V}\left(\widebar{\mathbf{H}}\right)\mathbf{s} + \mathbf{n} \nonumber \\[2mm]
&= \mathbf{H}^{\dagger}\left(\mathbf{t}\right)\mathbf{C}\left(\mathbf{t},\mathbf{z}_L\right)\mathbf{V}\left(\widebar{\mathbf{H}}\right)\mathbf{s} + \mathbf{n}, 
\end{align}
where $\mathbf{V}\left(\widebar{\mathbf{H}}\right)=\left[\mathbf{v}_1\left(\widebar{\mathbf{H}}\right),\dots,\mathbf{v}_K\left(\widebar{\mathbf{H}}\right)\right]\in\mathbb{C}^{M\times K}$ is defined as $\mathbf{V}\left(\widebar{\mathbf{H}}\right)=\frac{1}{f}\mathbf{C}^{-1}\left(\mathbf{t},\mathbf{z}_L\right)\mathbf{F}\left(\widebar{\mathbf{H}}\right)$, with $f = \left\|\mathbf{C}^{-1}\left(\mathbf{t},\mathbf{z}_L\right)\mathbf{F}\left(\widebar{\mathbf{H}}\right)\right\|/\sqrt{P_{\max}}$ denoting the scaling factor for ensuring that $\sum_{k\in\mathcal{K}}\left\|\mathbf{v}_k\left(\widebar{\mathbf{h}}_k\right)\right\|^2 \leq P_{\max}$. Note that we introduced mutual coupling in the expression of the precoder. Substituting the expression of $\mathbf{V}\left(\widebar{\mathbf{H}}\right)$ in Eq.~\eqref{eq:AA4}, we obtain
\begin{equation}\label{eq:AA5}
\mathbf{y} = \frac{1}{f}\mathbf{H}^{\dagger}\left(\mathbf{t}\right)\mathbf{F}\left(\widebar{\mathbf{H}}\right)\mathbf{s} + \mathbf{n}.
\end{equation}

\subsection{SINR and Sum-Rate}\label{subsec:2.3}
After RAPS, the signal-to-interference-plus-noise-ratio (SINR) of the $k$-th user for the given array configuration $\mathbf{t}$ is expressed in the case of a MAMP array as
\begin{equation}\label{eq:extra1}
\gamma_k\left(\mathbf{t}\right) = \frac{\left|\mathbf{h}_k^{\dagger}\left(\mathbf{t}\right)\mathbf{w}_k\right|^2}{\sum\limits_{i\in\mathcal{K}\setminus\{k\}}\left|\mathbf{h}_k^{\dagger}\left(\mathbf{t}\right)\mathbf{w}_i\right|^2 + \sigma^2},
\end{equation}
and in the case of an all-active array with tunable loads as
\begin{equation}\label{eq:extraAA}
\gamma_k\left(\mathbf{t}\right) = \frac{\left|\mathbf{h}_k^{\dagger}\left(\mathbf{t}\right)\mathbf{f}_k\right|^2}{\sum\limits_{i\in\mathcal{K}\setminus\{k\}}\left|\mathbf{h}_k^{\dagger}\left(\mathbf{t}\right)\mathbf{f}_i\right|^2 + f^2\sigma^2}.
\end{equation}
Notice that in~\eqref{eq:extraAA}, $f$ acts as a noise amplification factor. 

Under Gaussian signalling, the achievable rate of the $k$-th user and the sum-rate are given (in bps/Hz) by $R_k\left(\mathbf{t}\right)=\log_2\left(1+\gamma_k\left(\mathbf{t}\right)\right)$ and $R\left(\mathbf{t}\right)=\sum_{k\in\mathcal{K}}R_k\left(\mathbf{t}\right)$, respectively.

\section{RAPS: Problem Formulation and Solution}\label{sec:3}
Mutual coupling depends on the RA positions and load impedances. Hence, to simplify optimization, we decouple the respective tasks by performing RAPS based only on spatial correlation and relying solely on load impedances adjustment for controlling mutual coupling, given the selected ports\footnote{In this work, we assume a fixed active-passive partition grouping. Optimizing the active subset for mutual coupling concerns is an interesting extension that adds another combinatorial layer and is left for future consideration.}.

Given a heuristic linear precoder $\mathbf{F}$, we aim at optimizing the RAPS matrix to maximize the sum-rate. Under the unique array structure with $N$ ports dedicated to each RA, this problem is formulated, in the case of a MAMP array, as
\begin{subequations}\label{eq:22}
\begin{alignat}{2}
&&&\text{(P1): }\underset{\bm{\Sigma}}{\max} \ \sum_{k\in\mathcal{K}}\log_2\left(1+\frac{\left|\widehat{\mathbf{h}}_k^{\dagger}\left(\mathbf{p}\right)\bm{\Sigma}\widehat{\mathbf{w}}_k\right|^2}
{\sum\limits_{i\in\mathcal{K}\setminus\{k\}}\left|\widehat{\mathbf{h}}_k^{\dagger}\left(\mathbf{p}\right)\bm{\Sigma}\widehat{\mathbf{w}}_i\right|^2 + \sigma^2}\right) \label{eq:22a} \\
&&&\text{s.t.} \ \ \ \ b_{m,n}\in\{0,1\}, \quad \forall m\in\mathcal{M}, \ \forall n\in\mathcal{N}, \label{eq:22b} \\
&&& \ \ \ \ \ \ \ \sum_{n\in\mathcal{N}}b_{m,n} = 1, \quad \forall m\in\mathcal{M}, \label{eq:22c}
\end{alignat}
\end{subequations}
where $\widehat{\mathbf{w}}_k\in\mathbb{C}^L$ is the precoding vector for the $k$-th user adapted to $\widehat{\mathbf{H}}^{\dagger}\left(\mathbf{p}\right)$. Since we ignore mutual coupling during RAPS, as explained earlier, we solve the same exact problem for the all-active array with tunable loads scenario as well and adapt the precoder to its mutual coupling-dependent form during the optimization of the loading values.

(P1) is a combinatorial NP-hard problem with binary selection variables. The optimal exhaustive search (ES) approach has prohibitive complexity $N^M$. Thus, we apply meta-heuristics, and develop greedy algorithms and heuristic alternatives to effectively tackle this challenging optimization task. Regarding the former, we opt for BPSO and TS because they have empirically-known parameter values, handle efficiently large search spaces, and converge to high-quality solutions.

Note that the very first moment that the system becomes operational, the load impedances (and source voltages in case of a MAMP array) are optimized based on the given (initial) array configuration using the techniques described in Sec.~\ref{sec:4}. Next, RAPS is performed for the first time, and the loads (and source voltages) are optimized for the new RA positions, to terminate this first decoupled optimization cycle.

\begin{figure}[!t]
\begin{algorithm}[H]
\centering \small
\begin{algorithmic}[1]
\State Set the number of particles $P\!=\!30$, maximum number of iterations $T_{\max} = 200$, cognitive and social coefficients $c_1\!=\!c_2\!=\!2$, iteration index $t=0$, inertia weight $w^{(0)} = 0.9$, convergence tolerance $\varepsilon = 10^{-4}$.
\For{each particle $i \in \mathcal{P}$}
    \State Initialize $\mathbf{B}_i^{(0)}$ randomly such that $\sum_{n=1}^N B_{m,n}(i,0)\! =\! 1$, $\forall m \in \mathcal{M}$
    \State Initialize $\mathbf{V}_i^{(0)} \in \mathbb{R}^{M\times N}$ with random values in $[-1,1]$
    \State Set $\mathbf{P}_i^{(0)} = \mathbf{B}_i^{(0)}$
    \State Calculate $f\left(\mathbf{P}_i^{(0)}\right)$ as the sum-rate for configuration $\mathbf{P}_i^{(0)}$
\EndFor
\State Set $\mathbf{G}^{(0)} = \arg \ \max\limits_{\mathbf{P}_i^{(0)}} \ f\left(\mathbf{P}_i^{(0)}\right)$
\While{$\left|f\left(\mathbf{G}^{(t)}\right) - f\left(\mathbf{G}^{(t-1)}\right)\right| > \varepsilon$ \&\& $t < T_{\max}$}
    \For{each particle $i \in \mathcal{P}$}
        \State Calculate $f\left(\mathbf{B}_i^{(t)}\right)$ as the sum-rate for configuration $\mathbf{B}_i^{(t)}$
        \If{$f\left(\mathbf{B}_i^{(t)}\right) > f\left(\mathbf{P}_i^{(t-1)}\right)$}
            \State Update $\mathbf{P}_i^{(t)} = \mathbf{B}_i^{(t)}$
        \EndIf
        \If{$f\left(\mathbf{P}_i^{(t)}\right) > f\left(\mathbf{G}^{(t-1)}\right)$}
            \State Update $\mathbf{G}^{(t)} = \mathbf{P}_i^{(t)}$
        \EndIf
        \State Update velocity: 
        \begin{align}\label{eq:23}
\mathbf{V}_i^{(t)} =& w^{(t)}\mathbf{V}_i^{(t-1)} + c_1 r_1^{(t)}\left(\mathbf{P}_i^{(t)} - \mathbf{B}_i^{(t)}\right)\nonumber \\
&+ c_2 r_2^{(t)}\left(\mathbf{G}^{(t)} - \mathbf{B}_i^{(t)}\right)
\end{align}

        \For{each $m \in \mathcal{M}$}
            \State $S\left(V_{m,n}(i,t)\right) = \frac{1}{1 + e^{-V_{m,n}(i,t)}}$, $\forall n \in \mathcal{N}$
            \State Find $n^{\star} = \arg \ \max_n \ S\left(V_{m,n}(i,t)\right)$
            \State Set $B_{m,n^{\star}}(i,t) = 1$ and $B_{m,n}(i,t) = 0$ $\forall n \neq n^{\star}$
        \EndFor
        \State Update $t = t+1$
        \State Update $w^{(t)} = 0.9 - (0.5/T)t$
    \EndFor
\EndWhile
\State Set $\mathbf{B}^{\star} = \mathbf{G}^{(t)}$
\end{algorithmic}
\caption{BPSO for RAs Position Selection}
\label{algo:1}
\end{algorithm}
\vspace{-2em}
\end{figure}

\subsection{Binary Particle Swarm Optimization}\label{subsec:3.1}
The proposed BPSO algorithm (Alg.~\ref{algo:1}) uses a swarm of $P$ particles to explore the $N^M$ search space $\mathcal{T}$. Each particle $i\in\mathcal{P}\triangleq\{1,\dots,P\}$ represents a candidate solution (position matrix) $\mathbf{B}_i\triangleq\left[\mathbf{b}_1^T(i);\dots;\mathbf{b}_M^T(i)\right]\in\mathbb{B}^{M\times N}$, which encodes an array configuration $\mathbf{t}(i)\in\mathcal{T}$ and is evaluated using sum-rate as a fitness function $f\left(\cdot\right)$. In each iteration $t$, the personal best position of each particle $i$, $\mathbf{P}_i^{(t)}$, and the global best position of the swarm, $\mathbf{G}^{(t)}$, are updated based on particles' fitness. Particle velocity $\mathbf{V}_i^{(t)}$ is updated per Eq.~\eqref{eq:23}, using an inertia weight $w^{(t)}$, which controls the influence of $\mathbf{V}_i^{(t-1)}$ and decreases from $0.9$ to $0.4$ over iterations to balance exploration and exploitation, as well as cognitive and social constants $c_1=c_2=2$ respectively incorporating the personal and global best influence on velocity update. Position updates use a sigmoid function to map velocity matrix entries to probabilities and set only the element corresponding to the highest sigmoid probability in each row of the position matrix to $1$ while the others remain $0$, as per Eq.~\eqref{eq:22c}. The algorithm terminates upon convergence or after $T_{\max}$ iterations, yielding $\mathbf{B}^{\star}$. Worst-case complexity is $\mathcal{O}\left(T_{\max}PMNK\right)$, although in practice BPSO typically converges much faster.

\begin{figure}[!t]
\begin{algorithm}[H]
\centering \small
\begin{algorithmic}[1]
\State Set the maximum number of iterations $T_{\max}$, tabu tenure $\tau=\sqrt{MN}$, intensification period $T_i = M$, diversification period $T_d = 2M+1$, maximum number of consecutive non-improving iterations $I_c = M\log_2(N)$, non-improving iterations counter $i_c = 0$, iteration index $t=0$
\State Initialize $\mathbf{B}^{(0)}$ randomly such that $\sum_{n=1}^N B_{m,n}(0)\! =\! 1$, $\forall m \in \mathcal{M}$
\State Calculate $f(\mathbf{B}^{(0)})$ as the sum-rate for configuration $\mathbf{B}^{(0)}$
\State Set best solution $\mathbf{B}^{\star} = \mathbf{B}^{(0)}$ and $f^{\star} = f\left(\mathbf{B}^{(0)}\right)$
\State Initialize empty tabu list $\mathbb{T}=\emptyset$
\While{$t < T_{\max}$ \&\& $i_c < I_c$}
\State Set neighborhood:
\Statex $\mathcal{N}\left(\mathbf{B}^{(t)}\right) = \{\mathbf{B}': \exists! \ m\in\mathcal{M}, \exists (n_1,n_2)\in\mathcal{N}^2: b'_{m,n_1} = 1 - b^{(t)}_{m,n_1}, b'_{m,n_2} = 1 - b^{(t)}_{m,n_2}, b'_{m,n} = b^{(t)}_{m,n} \ \forall n\in\mathcal{N}\setminus\{n_1,n_2\}\}$
    \State Find best non-tabu neighbor $\mathbf{B}^{\prime}$ or tabu neighbor satisfying aspiration criterion: $\mathbf{B}^{\prime} = \arg\max\limits_{\mathbf{B}\in\mathcal{N}\left(\mathbf{B}^{(t)}\right)} \left\{f\left(\mathbf{B}\right): \left(m,n_{\text{old}},n_{\text{new}}\right)\notin\mathbb{T} \text{ or } f\left(\mathbf{B}\right) > f^{\star}\right\}$
    \If{$f\left(\mathbf{B}^{\prime}\right) > f^{\star}$}
        \State Update $\mathbf{B}^{\star} = \mathbf{B}^{\prime}$ and $f^{\star} = f\left(\mathbf{B}^{\prime}\right)$
        \State Reset $i_c = 0$
    \Else
        \State Increment $i_c = i_c + 1$  
    \EndIf
    \State Update $\mathbf{B}^{(t+1)} = \mathbf{B}^{\prime}$
    \State  Update $\mathbb{T} = \mathbb{T} \cup \{(m^{\star},n_{\text{old}}^{\star},n_{\text{new}}^{
    \star})\}$ where $(m^{\star},n_{\text{old}}^{\star},n_{\text{new}}^{\star}) = \Delta(\mathbf{B}^{\prime},\mathbf{B}^{(t)})$, $\Delta(\cdot,\cdot)$ returns the tuple describing the move between two solutions
    \If{$\left|\mathbb{T}\right| > \tau$}
        \State Remove oldest move from $\mathbb{T}$: $\mathbb{T} = \left\{(m,n_{\text{old}},n_{\text{new}})_i\right\}_{i=2}^{\left|\mathcal{T}\right|}$
    \EndIf
    \If{$t \bmod T_i = 0$}
    \State Set $\mathbf{B}^{(t+1)} = \mathbf{B}^{\star}$
    \State Update neighborhood: 
    \Statex $\mathcal{N}\left(\mathbf{B}^{(t+1)}\right) = \left\{\mathbf{B}^{\prime}: \sum_{m\in\mathcal{M}}\left\|\mathbf{b}^{\prime}_m - \mathbf{b}_m^{(t+1)}\right\|_0 \leq 2\right\}$
\EndIf
\If{$t \bmod T_d = 0$}
    \State Select random subset $\mathcal{M}_d \subseteq \mathcal{M}$ with $\left|\mathcal{M}_d\right| = \lfloor M/4 \rfloor$
    \State For each $m \in \mathcal{M}_d$: $b_{m,n}^{(t+1)} = \begin{cases} 1, & n = n_m^{\text{new}} \\ 0, & \text{otherwise} \end{cases}$, 
    \Statex \hspace{0.72cm} where $n_m^{\text{new}} \sim \mathcal{U}\left(\mathcal{N}\setminus\{n_m^{\text{cur}}\}\right)$
    \State Set $\mathbb{T} = \emptyset$
\EndIf
    \State Update $t = t+1$
\EndWhile
\State \Return $\mathbf{B}^{\star}$
\end{algorithmic}
\caption{TS for RAs Position Selection}
\label{algo:2}
\end{algorithm}
\end{figure}

\subsection{Tabu Search}\label{subsec:3.2}
In Alg.~\ref{algo:2}, we present the TS meta-heuristic for efficient RAPS purposes. TS is a local search method that avoids local optima by allowing non-improving moves and prohibiting previously visited solutions via a tabu list $\mathbb{T}$. Each solution is a compressed BSM $\mathbf{B}$, similar to BPSO. The tabu list with tenure $\tau = \sqrt{MN}$, which refers to the number of iterations a recent solution or move is forbidden, stores recent moves to prevent cycling, and tabu moves are identified as tuples $(m, n_{\text{old}}, n_{\text{new}})$. From an initial solution, TS iteratively explores neighborhood solutions $\mathcal{N}(\mathbf{B})$ by altering one port selection per RA, moving to the best neighboring solution even if it is suboptimal, and updating the best-known solution. Tabu moves are bypassed only if they meet the aspiration criterion, i.e., if they yield a better solution than the current best.  Algorithm's exploration/exploitation balance relies on intensification (returning to the best-known solution) and diversification (modifying $W$ random ports) whose occurrence depends on specified intervals $T_i = M$ and $T_d = 2M+1$, respectively. TS terminates when reaching $T_{\max}$ iterations or when there has been $I_c=M\log_2N$ consecutive iterations without improvement, yielding $\mathbf{B}^{\star}$. Worst-case complexity is $\mathcal{O}\left(T_{\max}MN^3 K\right)$ (higher than BPSO's due to neighborhood exploration), but TS needs $< T_{\max}$ iterations in practice. We should note that as a swarm heuristic algorithm, BPSO may  exhibit premature convergence while TS is typically more robust due to its memory and diversification mechanisms.
\begin{figure}[!t]
\begin{algorithm}[H]
\centering \small
\begin{algorithmic}[1]
\State Initialize $\mathbf{t} \leftarrow \emptyset$, $R_{\text{best}} \leftarrow 0$
\For{$i = 1$ to $M_a$}  \Comment{For each active RA}
\State $x_{\text{best}} \leftarrow 0$, $\Delta R_{\text{best}} \leftarrow -\infty$, $\mathbf{x}_{\text{pas}}^{\text{best}} \leftarrow \emptyset$
\For{$n = 1$ to $N$}  \Comment{Try each port of current active RA}
\State $\mathbf{t}_{\text{tmp}} \leftarrow \mathbf{t}$
\State $\mathbf{t}_{\text{tmp}}(i) \leftarrow n$  \Comment{Select $n$-th port for active RA $i$}
\For{$[x_1,\! \ldots,\!x_Q] \in {1,\! \ldots,\!N}^Q$}  \Comment{Try all passive RA port combinations}
\State $\mathbf{t}_{\text{tmp}}\left(M_a{+}(i{-}1)Q{+}1:M_a{+}iQ\right) \leftarrow [x_1,\ldots,x_Q]$
\State $R_{\text{curr}} \leftarrow \text{SR}\left(\mathbf{H}\left(:,\mathbf{t}_{\text{tmp}}\right),P_{\max},\sigma^2\right)$
\If{$R_{\text{curr}} - R_{\text{best}} > \Delta R_{\text{best}}$}
\State $x_{\text{best}} \leftarrow n$, $\Delta R_{\text{best}} \leftarrow R_{\text{curr}} - R_{\text{best}}$, $R_{\text{best}} \leftarrow R_{\text{curr}}$
\State $\mathbf{x}_{\text{pas}}^{\text{best}} \leftarrow [x_1,\ldots,x_Q]$
\EndIf
\EndFor
\EndFor
\State $\mathbf{t}(i) \leftarrow x_{\text{best}}$  \Comment{Fix best port for active RA $i$}
\State $\mathbf{t}\left(M_a{+}(i{-}1)Q{+}1:M_a{+}iQ\right) \leftarrow \mathbf{x}_{\text{pas}}^{\text{best}}$  \Comment{Fix ports for its passive RAs}
\EndFor
\State \Return $\mathbf{t}^{\star}$
\end{algorithmic}
\caption{Greedy RAPS (MAMP Array)}
\label{algo:3}
\end{algorithm}
\vspace{-2em}
\end{figure}

\subsection{Sparse BPSO/TS}\label{subsec:3.3}
To reduce the complexity of RAPS (at the potential cost of a performance loss) in the case of a MAMP RA array, we propose heuristic sparse BPSO and TS variants where RAPS is performed only for the $M_a$ active RAs. After selecting the ports of these RAs, we select the $Q$ ports adjacent to each one of them as the positions of the respective parasitic RAs in the corresponding sub-array. Hence, we replace $M$ in the previous worst-case complexity expressions with $M_a = M-M_p$. This architecture with subarray partitioning promotes more efficient channel estimation protocols by estimating only a subset of ports per subarray while the remaining channels can be inferred via correlation-based reconstruction across closely-spaced ports. A detailed design and optimization of such schemes is left for future work.

\subsection{Greedy RAPS Algorithm}\label{subsec:3.4}
We introduce a greedy RAPS algorithm which, as shown in Alg.~\ref{algo:3}, chooses, in the case of a MAMP RA array, active RAs' ports one-by-one (and their $Q$ adjacent ports as passive RAs), as long as they result in higher sum-rate than the current best configuration. Its complexity is $\mathcal{O}\left(M_aMN^2K\right)$, rendering it an efficient alternative to BPSO and TS, especially for moderate $N$. In the case of an all-active RA array with tunable loads, the algorithm chooses the best port for each RA one-by-one, thus we ignore the passive RAs selection part of the algorithm, as shown in Alg.~\ref{algo:4}.

\section{Load Impedances and Source Voltages}\label{sec:4}

\subsection{MAMP RA Array}\label{subsec:4.1}
In the case of a MAMP RA array, after RAPS, the $M$ selected RAs are divided into $M_a$ subgroups of $Q+1$ RAs each ($Q$ passive and one active). If there is an even number of passive RAs, i.e., $Q = 2r$, $r\in\mathcal{N}$, then the $\left(\frac{Q}{2}+1\right)$-th RA in each sub-array is the active one; otherwise, that is, if $Q= 2r+1$, it is either the $\frac{Q}{2}$-th or the $\left(\frac{Q}{2}+1\right)$-th RA the one fed by the single RF chain (random coin toss).

To resemble an all-active array for given precoding and modulation schemes, it should be $\mathbf{i} = \mathbf{W}\mathbf{s}$, i.e., $i_m = \sum_{k\in\mathcal{K}}w_{m,k}s_k$, where $\mathbf{W}$ is adapted to $\widebar{\mathbf{H}}$. We can control the input currents $\mathbf{i}$ by adjusting the voltage sources and loading values (for a given array configuration $\mathbf{t}$ after RAPS). Specifically, by rearranging Eq.~\eqref{eq:4}, we obtain $\left(\mathbf{Z}\left(\mathbf{t}\right)+\mathbf{X}\right)\mathbf{i} = \mathbf{v}$, where $Z_{m,m}=z_A$, $X_m = z_0$ for $m\in\mathcal{M}_a$, and $v_m = 0$ for $m\in\mathcal{M}_p$. Combining these expressions, we obtain
\vspace{-1mm}
\begin{subequations}\label{eq:21}
\begin{alignat}{2}
v_m & = \sum_{k\in\mathcal{K}} \left(\sum_{j=1}^M Z_{m,j}w_{j,k} + z_0w_{m,k}\right)s_k, \quad m\in\mathcal{M}_a, \label{eq:21a} \\[-4mm]
X_m & = \frac{-\sum\limits_{k\in\mathcal{K}} \left(\sum\limits_{j=1}^M Z_{m,j}w_{j,k}\right)s_k}{\sum\limits_{k\in\mathcal{K}} w_{m,k}s_k}, \quad m\in\mathcal{M}_p. \label{eq:21b}
\vspace{-2mm}
\end{alignat}
\end{subequations}
We note that both the real (resistance) and imaginary (reactance) parts of the load impedances can take both positive and negative values. An implementation of a respective loading circuit based on CMOS transistors is given in~\cite{ESPARQAM}. To avoid the amplification noise attributed to negative resistance values, we replace them with their absolute value, with only moderate performance degradation, thus resulting in a compelling trade-off. This is because the magnitude typically dominates the resulting current distribution and radiation behavior more strongly than the sign of the real part, especially when the former is large. As such, the magnitude-dependent coupling effects remain largely intact, preserving most of the desired pattern-shaping behavior. Also, with large enough self-impedances, we ensure that the array radiates instead of consuming power~\cite{ESPARQuantLoads}.
\begin{figure}[!t]
\begin{algorithm}[H]
\centering \small
\begin{algorithmic}[1]
\State Initialize $\mathbf{t} \leftarrow \emptyset$, $R_{\text{best}} \leftarrow 0$

\For{$i = 1$ to $M$}  \Comment{For each RA}
    \State $x_{\text{best}} \leftarrow 0$, $\Delta R_{\text{best}} \leftarrow -\infty$
    
    \For{$n = 1$ to $N$}  \Comment{Try each port of current RA}
        \State $\mathbf{t}_{\text{tmp}} \leftarrow \mathbf{t}$
        \State $\mathbf{t}_{\text{tmp}}(i) \leftarrow n$  \Comment{Select $n$-th port for RA $i$}
        
        \State $R_{\text{curr}} \leftarrow \text{SR}\left(\mathbf{H}\left(:,\mathbf{t}_{\text{tmp}}\right),P_{\max},\sigma^2\right)$
        
        \If{$R_{\text{curr}} - R_{\text{best}} > \Delta R_{\text{best}}$}
            \State $x_{\text{best}} \leftarrow n$
            \State $\Delta R_{\text{best}} \leftarrow R_{\text{curr}} - R_{\text{best}}$
            \State $R_{\text{best}} \leftarrow R_{\text{curr}}$
        \EndIf
    \EndFor
    
    \State $\mathbf{t}(i) \leftarrow x_{\text{best}}$  \Comment{Fix best port for RA $i$}
\EndFor

\State \Return $\mathbf{t}^{\star}$
\end{algorithmic}
\caption{Greedy RAPS (All-Active Array)}
\label{algo:4}
\end{algorithm}
\vspace{-2em}
\end{figure}

\subsection{All-Active RA Array With Tunable Loads}\label{subsec:4.2}
In the case of an all-active RA array with tunable loads, we seek to optimize the loading values after RAPS to maximize the sum-rate for the given array configuration. This problem is formulated as follows
\begin{subequations}\label{eq:Opt}
\begin{alignat}{2}
&&&\text{(P2): }\underset{\mathbf{z}_L}{\max} \ \sum_{k\in\mathcal{K}}\log_2\left(1+\frac{\left|\mathbf{h}_k^{\dagger}\left(\mathbf{t}\right)\mathbf{f}_k\left(\mathbf{H}\right)\right|^2}{\sum\limits_{i\in\mathcal{K}\setminus\{k\}}\left|\mathbf{h}_k^{\dagger}\left(\mathbf{t}\right)\mathbf{f}_i\left(\mathbf{H}\right)\right|^2 +\! f^2\sigma^2}\right) \label{eq:Opta} \\[-2mm]
&&&\text{s.t.} \ \ \ \ f = \frac{1}{\sqrt{P_{\max}}}\left\|\mathbf{C}^{-1}\left(\mathbf{t},\mathbf{z}_L\right)\mathbf{F}\left(\widebar{\mathbf{H}}\right)\right\|. \label{eq:Optb}
\end{alignat}
\end{subequations}
Since the only variable is $f$, and the channels $\mathbf{h}_{k}^{\dagger}\left(\mathbf{t}\right)$ for a given array configuration $\mathbf{t}$ do not depend on mutual coupling (see Eq.~\eqref{eq:AA5}), and thus on the loading values, this problem simplifies to
\begin{equation}\label{eq:Opt2}
\text{(P3): }\underset{\mathbf{z}_L}{\min}\left\|\mathbf{C}^{-1}\left(\mathbf{t},\mathbf{z}_L\right)\mathbf{F}\left(\widebar{\mathbf{H}}\right)\right\|.
\end{equation}
This unconstrained minimization problem can be solved using complex gradient descent. By expressing ${\mathbf{C}}^{-1}({\mathbf{t}}, {\mathbf{z}}_L)$ explicitly using Eq.~\eqref{eq:7b}, we obtain
\begin{equation}\label{eq:Opt3}
{\mathbf{C}}^{-1}({\mathbf{t}}, {\mathbf{z}}_L) = [{\mathbf{Z}}_T({\mathbf{t}}, {\mathbf{z}}_L)]^{-1}(z_A{\mathbf{I}}_M + {\mathbf{X}})^{-1}.
\end{equation}
Given that ${\mathbf{X}} = \text{diag}({\mathbf{z}}_L)$, we can write
\begin{equation}\label{eq:Opt4}
(z_A{\mathbf{I}}_M + {\mathbf{X}})^{-1} = \text{diag}([1/(z_A + z_1), \ldots, 1/(z_A + z_M)])
\end{equation}
Then, the gradient with respect to ${\mathbf{z}}_L$ can be computed as
\begin{subequations}\label{eq:Opt5}
\begin{alignat}{2}
&\nabla_{{\mathbf{z}}_L}\|{\mathbf{C}}^{-1}{\mathbf{F}}\| = -\frac{\operatorname{diag}\left(\mathbf{U}\right)^{\dagger}}{2\left\|\mathbf{C}^{-1}\mathbf{F}\right\|}, \label{eq:Opt5a} \\
&\mathbf{U} = \left(z_A{\mathbf{I}}_M + {\mathbf{X}}\right)^{-1} \cdot [{\mathbf{Z}}_T]^{-1}{\mathbf{F}} \cdot {\mathbf{F}}^H[{\mathbf{Z}}_T]^{-1}. \label{eq:Opt5b}
\end{alignat}
\end{subequations}
The iterative update rule is given by
\begin{equation}\label{eq:Opt6}
{\mathbf{z}}_L^{(n+1)} = {\mathbf{z}}_L^{(n)} - \mu\nabla_{{\mathbf{z}}_L}\|{\mathbf{C}}^{-1}{\mathbf{F}}\|,
\end{equation}
where $n$ is the iteration index and $\mu$ is the step size that can be determined through line search. The algorithm converges when the norm of the gradient falls below a predetermined threshold $\epsilon$ or after reaching a maximum number of iterations. The computational complexity of the proposed gradient descent algorithm is $O(TM^3)$, where $T$ is the number of iterations until convergence. This complexity is primarily attributed to the matrix inversion operations required to compute ${\mathbf{C}}^{-1}$ in each iteration. While this represents a significant computational load for large arrays, the optimization needs to be performed only when the array configuration changes substantially, making it feasible for practical implementations. Thus, the proposed solution efficiently minimizes the effect of mutual coupling by optimizing the loading values, while maintaining computational tractability.

\section{Performance Analysis and Robust Designs With Quantized Loads}\label{sec:5}
In this section, we analyze a more realistic array model by accounting for imperfections in electronic components, where only quantized load values with finite precision are practically achievable~\cite{TunLoads}. For an all-active RA array, the quantized load values of each tunable impedance can be expressed as
\begin{align}\label{eq:tunImpErr}
\hat{X}_m = X_m + e_m, \text{ for } m=1,2,\ldots,M,
\end{align}
where $\hat{X}_m$ is the quantized load value for the $m$-th RA, and $e_m$ is the quantization error, modeled as a complex Gaussian random variable with zero mean and variance $\epsilon$, i.e., $e_m \sim \mathcal{CN}(0,\epsilon)$. For a MAMP array, the load index takes values $m=M_{a+1}, \dots,M$. By assuming a quantization level $D$, the value of each tunable load is expressed as
\begin{align}\label{eq:tunImpErr2}
\hat{X}_m = \mu_m D + j\nu_m D,
\end{align}
where $ \mu_m, \nu_m\in \left\{ 0, \pm 1, \pm 2, \ldots \right\}$. 

\subsection{MAMP RA Array}\label{subsec:5.1}
From Eq. \eqref{eq:4} the current vector with quantized loads is
\begin{align}\label{eq:currentVecQuant}
\hat{{\bf{i}}}\left({\bf{t}}, {\bf{z}}_L\right) &= \hat{{\bf{Z}}}_T\left({\bf{t}}, {\bf{z}}_L\right) {\bf{v}} \nonumber \\
&= \left( {\bf{Z}}\left({\bf{t}}\right) + {\bf{X}} + {\bf{E}}_{q} \right)^{-1} {\bf{v}} \nonumber \\
&= \left( {\bf{Z}}_T^{-1}\left({\bf{t}}, {\bf{z}}_L\right) + {\bf{E}}_{q} \right)^{-1} {\bf{v}},
\end{align}
where ${\bf{E}}_q \triangleq {\rm{diag}}({\bf{e}}_q) \in \mathbb{C}^{M \times M}$ is the impedance error matrix, and ${\bf{e}}_q = [0,\ldots,0,e_{M_a+1},\ldots,e_{M}]^T \in \mathbb{C}^M$ represents the error in the load value. We assume that the value of load errors is statistically independent to the impedance value. In cases with impedance errors, the feeding voltages remain the same as given in Eq.~\eqref{eq:21a}, whereas the loading values are computed by manipulating Eq.~\eqref{eq:currentVecQuant} and are given by
\begin{align}\label{eq:loadValuesQuant}
\hat{X}_m & = -\left[\frac{\sum\limits_{k\in\mathcal{K}} \left(\sum\limits_{j=1}^M Z_{m,j}w_{j,k}\right)s_k}{\sum\limits_{k\in\mathcal{K}} w_{m,k}s_k} + e_m\right], \quad m\in\mathcal{M}_p.
\vspace{-1mm}
\end{align}
Since the feeding voltages remain the same, we can  relate the curent vector with quantized loads with that of the ideal current vector as
\begin{align}
\left( {\bf{Z}}_T^{-1}\left({\bf{t}}, {\bf{z}}_L\right) + {\bf{E}}_{q} \right) \hat{{\bf{i}}}\left({\bf{t}}, {\bf{z}}_L\right) = {\bf{Z}}_T^{-1}\left({\bf{t}}, {\bf{z}}_L\right) {\bf{i}}\left({\bf{t}}, {\bf{z}}_L\right),
\end{align}
which, after a few algebraic manipulations, is given by 
\begin{align}\label{eq:currentVecErrors}
\hat{{\bf{i}}}\left({\bf{t}}, {\bf{z}}_L\right) &= {\bf{i}}\left({\bf{t}}, {\bf{z}}_L\right) - \left( {\bf{Z}}_T^{-1}\left({\bf{t}}, {\bf{z}}_L\right) + {\bf{E}}_{q} \right)^{-1} {\bf{E}}_{q} {\bf{i}}\left({\bf{t}}, {\bf{z}}_L\right).
\end{align}
The received signal vector at all user terminals while taking into account impedance errors is then expressed as
\begin{align}\label{eq:recSigQuant}
\mathbf{y} &= \widebar{\mathbf{H}}^{\dagger}\left(\mathbf{t},\mathbf{z}_L\right)\hat{\mathbf{i}} + \mathbf{n} \nonumber \\[1mm]
&= \widebar{\mathbf{H}}^{\dagger}\left(\mathbf{t},\mathbf{z}_L\right)\mathbf{i} + \mathbf{n} \nonumber \\
& \quad - \widebar{\mathbf{H}}^{\dagger}\left(\mathbf{t},\mathbf{z}_L\right) \left( {\bf{Z}}_T^{-1}\left({\bf{t}}, {\bf{z}}_L\right) + {\bf{E}}_{q} \right)^{-1} {\bf{E}}_{q} {\bf{i}}\left({\bf{t}}, {\bf{z}}_L\right) \nonumber \\
&= \widebar{\mathbf{H}}^{\dagger}\left(\mathbf{t},\mathbf{z}_L\right)\hat{\mathbf{i}} + \mathbf{n} + \mathbf{n}_q,
\end{align}
where $\mathbf{n}_q = -\widebar{\mathbf{H}}^{\dagger}\left(\mathbf{t},\mathbf{z}_L\right) \left( {\bf{Z}}_T^{-1}\left({\bf{t}}, {\bf{z}}_L\right) + {\bf{E}}_{q} \right)^{-1} {\bf{E}}_{q} {\bf{i}}\left({\bf{t}}, {\bf{z}}_L\right)$ represents an additive noise term that is introduced by the quantized loads. For each user $k$, the variance of the added noise due to quantization is $\sigma_q^2 = M \xi \epsilon $, where $\xi=\left\lVert \left( {\bf{Z}}_T^{-1}\left({\bf{t}}, {\bf{z}}_L\right) + {\bf{E}}_{q} \right)^{-1} \right\rVert ^2$. Therefore, the SINR of the $k$-th user, in the presence of impedance errors, is given by
\begin{align}\label{eq:quantSINR}
\gamma_k\left(\mathbf{t}\right) = \frac{\left|\mathbf{h}_k^{\dagger}\left(\mathbf{t}\right)\mathbf{w}_k\left(\mathbf{H}\right)\right|^2}{\sum\limits_{i\in\mathcal{K}\setminus\{k\}}\left|\mathbf{h}_k^{\dagger}\left(\mathbf{t}\right)\mathbf{w}_i\left(\mathbf{H}\right)\right|^2 + \sigma^2 + \sigma_q^2}.
\end{align}

\begin{figure*}[!t]
\begin{minipage}[b]{.324\linewidth}
	\centering
	\includegraphics[width=\columnwidth]{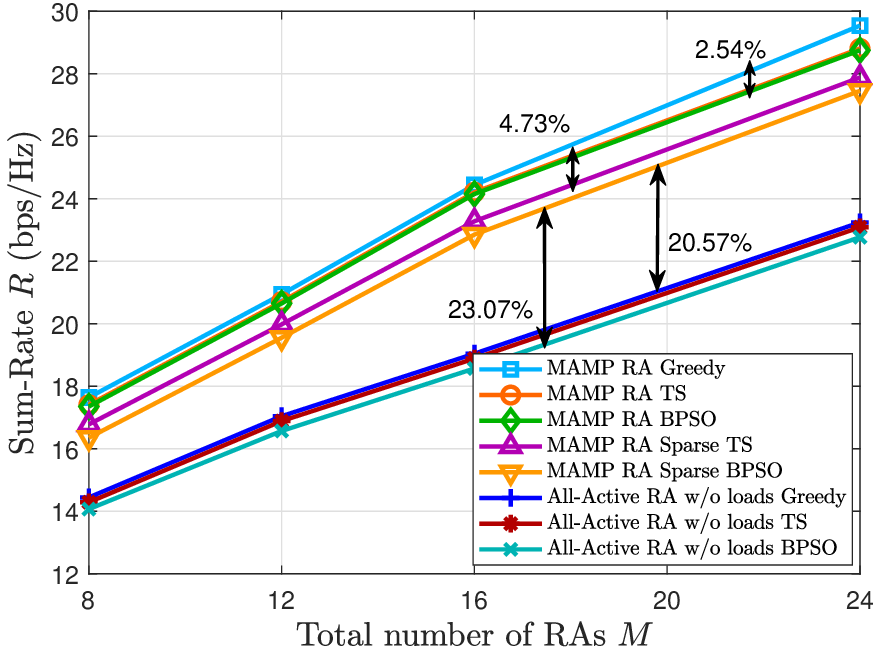}
	\caption{SR vs. number of RAs: MAMP vs. all-active RAs without tunable loads.}
	\label{fig:4}
	\vspace{-2mm}
\end{minipage}
\begin{minipage}[b]{.324\linewidth}
	\centering
	\includegraphics[width=\columnwidth]{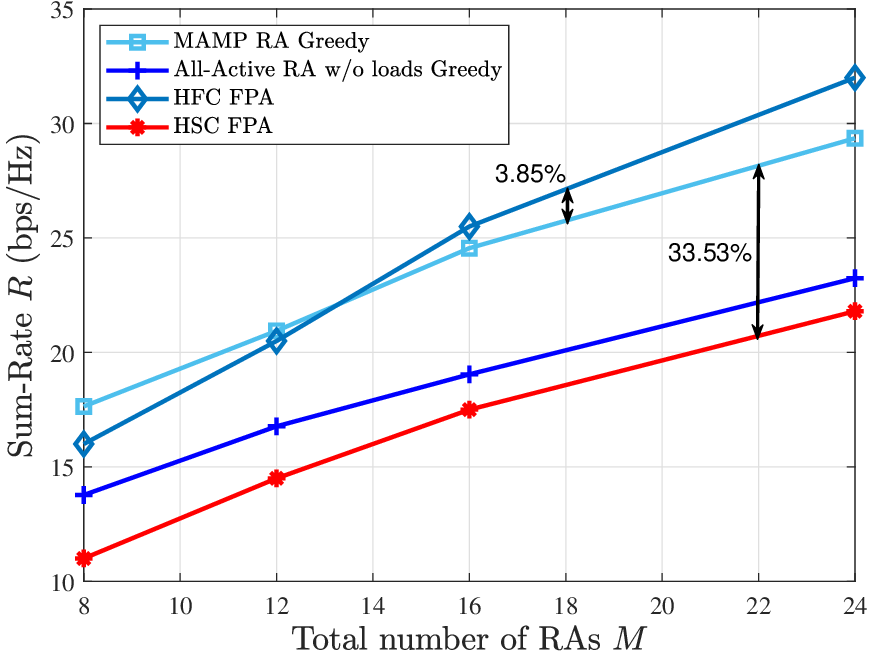}
	\caption{SR vs. number of RAs: MAMP RA vs. HFC and HSC FPA arrays.}
	\label{fig:5}
	\vspace{-2mm}
\end{minipage}
	\hspace{0.5mm}
\begin{minipage}[b]{.324\linewidth}
	\centering
	\includegraphics[width=\columnwidth]{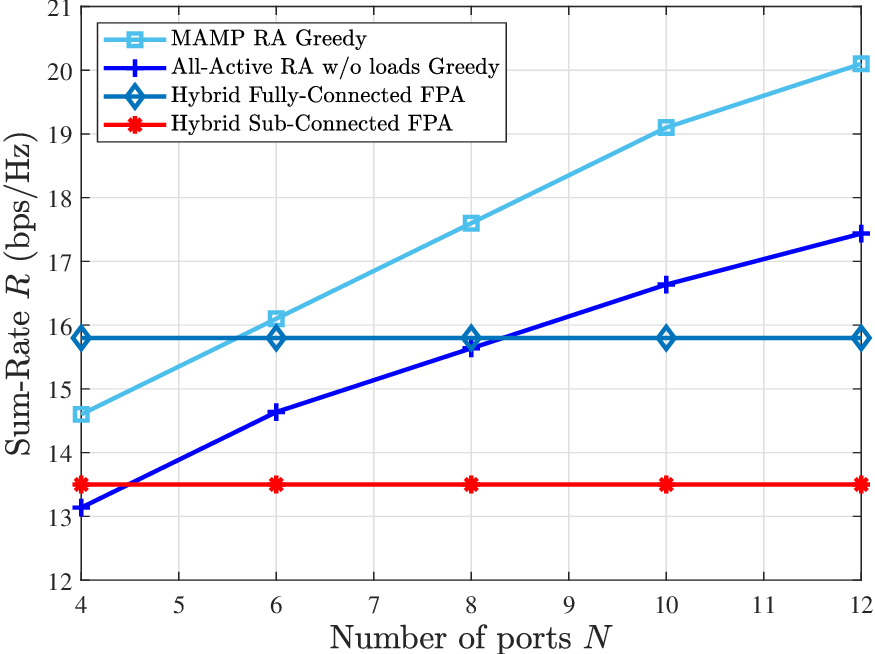}
	\caption{SR vs. number of ports for the MAMP array.}
	\label{fig:6}
	\vspace{-2mm}
\end{minipage}
\end{figure*}

Typically, a joint optimization of the feeding voltages and the quantized loads is applied to achieve an improved performance. More specifically, by considering a variation in the quantized loads and feeding voltages, the current vector with impedance errors is expressed as
\begin{align}\label{eq:currentVecTilde}
\tilde{{\bf{i}}}\left({\bf{t}}, {\bf{z}}_L\right) &= \left( {\bf{Z}}_T^{-1}\left({\bf{t}}, {\bf{z}}_L\right) + \tilde{{\bf{E}}}_{q} \right)^{-1} \tilde{{\bf{v}}}\nonumber \\
&= \left( {\bf{Z}}_T^{-1}\left({\bf{t}}, {\bf{z}}_L\right) + {\bf{E}}_{q} + D \bar{{\bf{E}}} \right)^{-1} \left( {\bf{v}} + \Delta {\bf{v}} \right),
\end{align}
where $\tilde{{\bf{v}}} = {\bf{v}} + \Delta {\bf{v}}$, and $\Delta {\bf{v}}$ represents the variation in the feeding voltages. Furthermore, $\tilde{{\bf{E}}}_{q} = {\bf{E}}_{q} + D \bar{{\bf{E}}}$, where $\bar{{\bf{E}}}$ is a diagonal matrix of size $M \times M$ (with each entry being a complex integer) that models the additional quantized load impedance values. However,  the variation in the quantized loads contributes to the increase in the noise power, which leads to an additional performance degradation \cite{ESPARQuantLoads}. Thus, the optimal case is to optimize the feeding voltages only and set $\bar{{\bf{E}}} = {\bf{0}}$. Accordingly, by using Eq.~\eqref{eq:4} and Eq.~\eqref{eq:currentVecTilde}, we obtain
\begin{align}
\tilde{{\bf{i}}}\left({\bf{t}}, {\bf{z}}_L\right) &= \left( {\bf{Z}}_T^{-1}\left({\bf{t}}, {\bf{z}}_L\right) + {\bf{E}}_{q} \right)^{-1} \Delta {\bf{v}}\nonumber \\
& \quad + \left( {\bf{Z}}_T^{-1}\left({\bf{t}}, {\bf{z}}_L\right) + {\bf{E}}_{q} \right)^{-1} {\bf{Z}}_T^{-1}\left({\bf{t}}, {\bf{z}}_L\right) {\bf{i}}\left({\bf{t}}, {\bf{z}}_L\right)\nonumber \\
&= {\bf{i}}\left({\bf{t}}, {\bf{z}}_L\right) + \left( {\bf{Z}}_T^{-1}\left({\bf{t}}, {\bf{z}}_L\right) + {\bf{E}}_{q} \right)^{-1} \Delta {\bf{v}} \nonumber \\
& \quad - \left( {\bf{Z}}_T^{-1}\left({\bf{t}}, {\bf{z}}_L\right) + {\bf{E}}_{q} \right)^{-1} {\bf{E}}_{q} {\bf{i}}\left({\bf{t}}, {\bf{z}}_L\right).
\end{align}
Subsequently, the difference between the current vector with impedance errors and the ideal current vector is given by
\begin{align}\label{eq:Delta_i}
\Delta {\bf{i}} &= \tilde{{\bf{i}}}\left({\bf{t}}, {\bf{z}}_L\right) - {\bf{i}}\left({\bf{t}}, {\bf{z}}_L\right)\nonumber \\[2mm]
&= \left( {\bf{Z}}_T^{-1}\left({\bf{t}}, {\bf{z}}_L\right) + {\bf{E}}_{q} \right)^{-1} \Delta {\bf{v}} \nonumber \\
& \quad - \left( {\bf{Z}}_T^{-1}\left({\bf{t}}, {\bf{z}}_L\right) + {\bf{E}}_{q} \right)^{-1} {\bf{E}}_{q} {\bf{i}}\left({\bf{t}}, {\bf{z}}_L\right)\nonumber \\[2mm]
&= {\bf{A}}^{-1} \Delta {\bf{v}} - {\bf{A}}^{-1} {\bf{E}}_{q} {\bf{i}}\left({\bf{t}}, {\bf{z}}_L\right),
\end{align}
where ${\bf{A}}$ is defined as ${\bf{A}} \triangleq {\bf{Z}}_T^{-1}\left({\bf{t}}, {\bf{z}}_L\right) + {\bf{E}}_{q}$.

To compensate for the performance loss induced by the presence of errors in the load values, we consider minimizing the error between the current vector with impedance errors and the ideal current vector. Based on Eq.~\eqref{eq:Delta_i}, we aim to find the optimal $\Delta {\bf{v}}$ that minimizes $\lVert \Delta {\bf{i}}\rVert ^2$. Thus, the optimization problem is formulated as
\begin{subequations}\label{eq:minDelta}
\begin{alignat}{2}
&&&\text{(P4): }\underset{\Delta {\bf{v}}}{\min} \quad \left\lVert {\bf{A}}^{-1} \Delta {\bf{v}} - {\bf{A}}^{-1} {\bf{E}}_{q} {\bf{i}}\left({\bf{t}}, {\bf{z}}_L\right)\right\rVert ^2 \label{eq:minDelta_a} \\[2mm]
&&&\text{s.t.} \ \ \ \ \Delta {\bf{v}} \geq {\bf{0}}, \quad \forall m\in\mathcal{M}, \label{eq:minDelta_b}
\end{alignat}
\end{subequations}
which can be efficiently solved using convex optimization tools, such as CVX~\cite{CVX}.

\subsection{All-Active RA Array With Tunable Loads}\label{subsec:5.2}
Under quantized loads, the mutual coupling matrix becomes
\begin{equation}
\hat{{\mathbf{C}}}({\mathbf{t}}, {\mathbf{z}}_L) = (z_A{\mathbf{I}}_M + \hat{{\mathbf{X}}}){\mathbf{Z}}_T({\mathbf{t}}, {\mathbf{z}}_L),
\end{equation}
where $\hat{{\mathbf{X}}} = {\mathbf{X}} + {\mathbf{E}}_q$ and ${\mathbf{E}}_q = \text{diag}({\mathbf{e}})$ is the error matrix, and the received signal vector with quantized loads becomes
\begin{equation}
\begin{aligned}
{\mathbf{y}} &= {\mathbf{H}}^{\dagger}({\mathbf{t}})\hat{{\mathbf{C}}}({\mathbf{t}}, {\mathbf{z}}_L){\mathbf{V}}({\mathbf{H}}){\mathbf{s}} + {\mathbf{n}} \\
&= \frac{1}{\hat{f}}{\mathbf{H}}^{\dagger}({\mathbf{t}}){\mathbf{F}}({\mathbf{H}}){\mathbf{s}} + {\mathbf{n}},
\end{aligned}
\end{equation}
where $\hat{f} = \|\hat{{\mathbf{C}}}^{-1}({\mathbf{t}}, {\mathbf{z}}_L){\mathbf{F}}({\mathbf{H}})\|/\sqrt{P_{\text{max}}}$ is the scaling factor under quantized loads. Therefore, the SINR for the $k$-th user is expressed as
\begin{equation}
\gamma_k({\mathbf{t}}) = \frac{|{\mathbf{h}}_k^{\dagger}({\mathbf{t}}){\mathbf{f}}_k({\mathbf{H}})|^2}{\sum_{i \in \mathcal{K}\setminus\{k\}} |{\mathbf{h}}_k^{\dagger}({\mathbf{t}}){\mathbf{f}}_i({\mathbf{H}})|^2 + \hat{f}^2\sigma^2}.
\end{equation}

To develop a robust design that minimizes the impact of quantization errors, we minimize the difference between the scaling factors under quantized and ideal loads subject to the quantization constraints
\begin{subequations}\label{eq:minf}
\begin{alignat}{2}
&&&\text{(P5): }\underset{\mathbf{z}_L}{\min} \quad |\hat{f} - f|^2 \label{eq:minf_a} \\
&&&\text{s.t.} \ \ \ \ {\mathbf{z}}_L = \mu{\mathbf{D}} + j\nu{\mathbf{D}}, \quad \mu,\nu \in \mathbb{Z}^M \label{eq:minf_b}
\end{alignat}
\end{subequations}
This mixed-integer optimization problem can be solved efficiently as follows:
\begin{enumerate}
\item Solve for continuous ${\mathbf{z}}_L$ using gradient descent on
\begin{equation}
\min_{{\mathbf{z}}_L} \left|\frac{\|\hat{{\mathbf{C}}}^{-1}({\mathbf{t}}, {\mathbf{z}}_L){\mathbf{F}}({\mathbf{H}})\|}{\sqrt{P_{\text{max}}}} - \frac{\|{\mathbf{C}}^{-1}({\mathbf{t}}, {\mathbf{z}}_L){\mathbf{F}}({\mathbf{H}})\|}{\sqrt{P_{\text{max}}}}\right|^2
\end{equation}
\item Quantize the solution to the nearest feasible load values.
\item Iterate until convergence or maximum iterations reached.
\end{enumerate}

The computational complexity is $O(TM^3)$, where $T$ is the number of gradient descent steps. This approach provides a practical solution that directly optimizes the precoding performance while accounting for quantization effects in the hardware implementation.

\section{Numerical Evaluations}\label{sec:6}

\subsection{MAMP RA Array}\label{subsec:6.1}
In this section, we comparatively evaluate via numerical Monte Carlo simulations in MATLAB the proposed designs for the MAMP RA array, in terms of the achievable sum-rate (SR), against the following benchmarks: an all-active array with $M_a$ RAs (without tunable loads) that employs the same RAPS schemes as well as HFC and HSC arrays with $M_a$ RF chains, $M$ FPAs, and, in the case of HSC, $M_a$ subarrays with $Q+1$ FPAs in each. Note that HFC and HSC FPA arrays require $M_a\times M$ and $M$ phase shifters, respectively, while the MAMP array only needs $M_p = M-M_a$ analog loads and few variable resistors~\cite{ESPARQAM}. Given that each phase shifter consumes 30 mW and has an insertion loss of 0.7 dB, while varactors have negligible power consumption~\cite{Heath2025} and PIN diodes consume around 10 mW, it becomes apparent that MAMP structures provide significant power savings and energy efficiency gains over digital and hybrid designs~\cite{Heath2025}.

Unless otherwise stated, the simulation parameters are $f_c = 28$ GHz, $\bar{d} = \lambda/16$, $K=2$, $N=4$, $N_c = 3$, and $N_p = 10$. Note that $\mathbf{i}= \alpha\mathbf{F}\mathbf{s}$ directly translates transmit symbols into antenna currents under Gaussian signaling; hence, we use Shannon formula as the MAMP RA array's SR upper bound, although QPSK modulation is employed, as in~\cite{ESPARQuantLoads}. We consider short-range mmWave access scenarios (e.g., indoor) since the proposed MAMP array has a massive number of ports but a limited number of physical RAs.

\begin{figure*}[!t]
\begin{minipage}[b]{.48\linewidth}
	\centering 
	\includegraphics[width=\columnwidth]{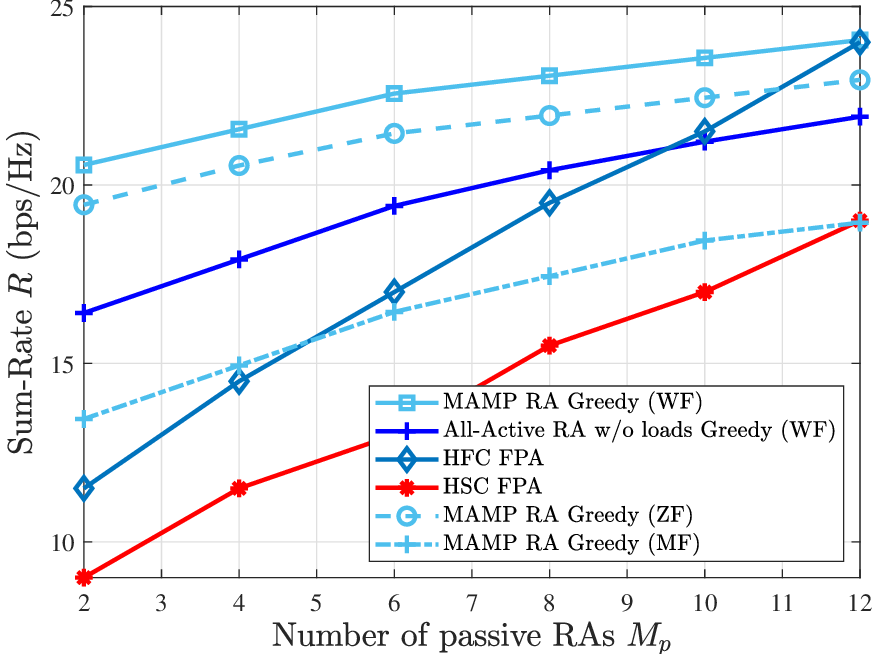}
	\caption{SR vs. number of passive antennas for MF, ZF, and WF precoding.}
	\label{fig:7}
\end{minipage}
\hspace{1mm}
\begin{minipage}[b]{.48\linewidth}
	\centering 
	\includegraphics[width=\columnwidth]{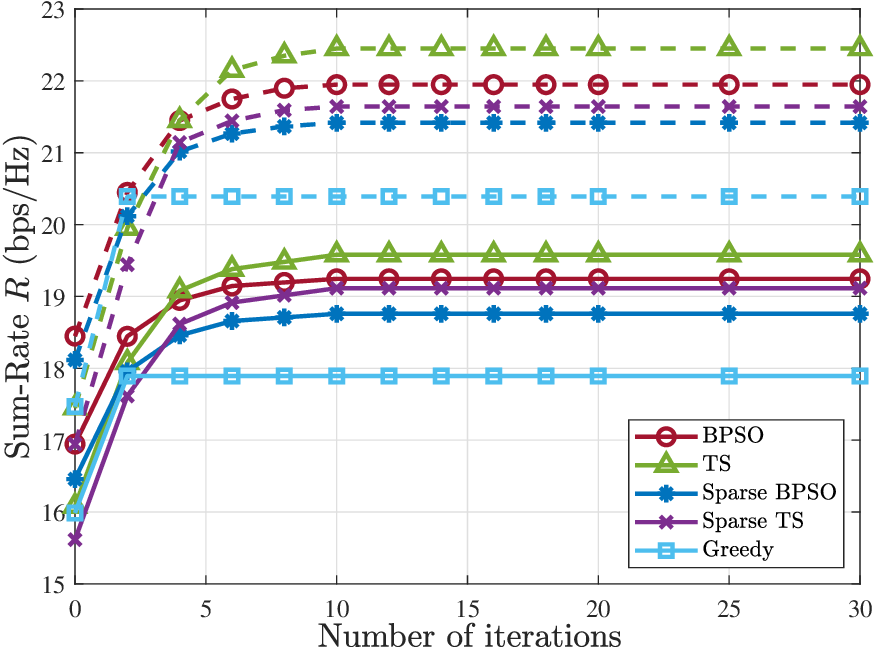}
	\caption{Convergence behavior of proposed RAPS algorithms for $N=6$ (solid lines) and $N=10$ (dashed lines).}
	\label{fig:8}
\end{minipage}
\end{figure*}

\begin{figure*}
\centering
\begin{subfigure}[t]{0.32\textwidth}
	\centering
	\includegraphics[width=\linewidth]{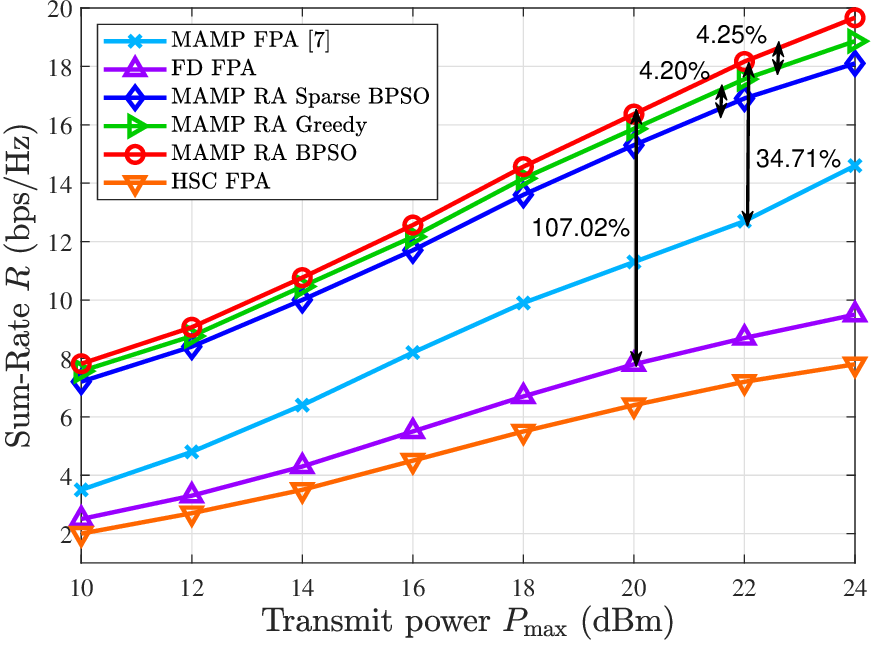}
	\caption{}
	\label{fig:9a}
\end{subfigure}%
~
\begin{subfigure}[t]{0.32\textwidth}
	\centering
	\includegraphics[width=\linewidth]{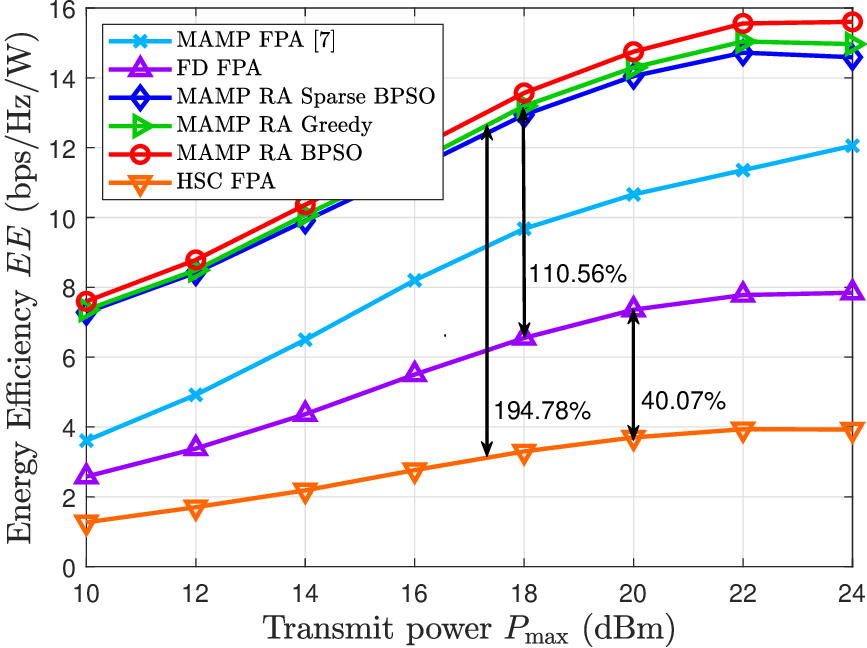}
	\caption{}
	\label{fig:9b}
\end{subfigure}
~
\begin{subfigure}[t]{0.32\textwidth}
	\centering
	\includegraphics[width=\linewidth]{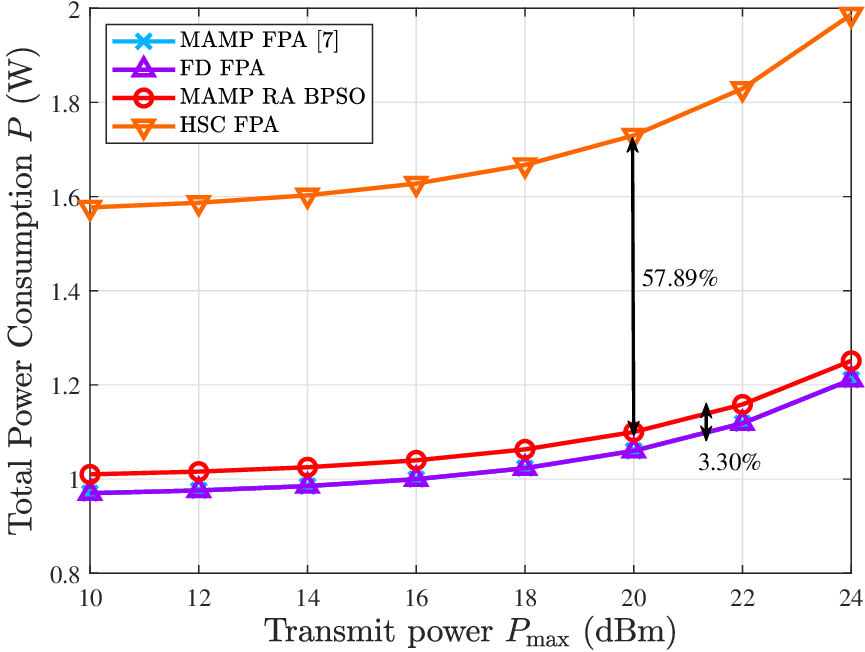}
	\caption{}
	\label{fig:9c}
\end{subfigure}
\caption{Performance of the MAMP RA array versus transmit power: (a) SR. (b) Energy efficiency. (c) Power consumption.}
\label{fig:extra}
\end{figure*}

In Figs.~\ref{fig:4} and~\ref{fig:5}, we plot the SR vs. the total number of RAs $M = [8; 12; 16; 24]$, corresponding to $M_a = [4; 4; 8; 8]$ active and $M_p = [4; 8; 8; 16]$ passive RAs, under the use of WF precoding. In Fig.~\ref{fig:4}, we see that the MAMP RA array with sparse BPSO/TS RAPS achieves about 21-23\% higher SR than the all-active RA array without loads, attributed to the additional ports and passive antennas. Non-sparse variants further improve SR by about 4.7\%, with TS marginally outperforming BPSO in both cases. Greedy RAPS provides an additional 2.5\% gain. SR improves with increasing $M$ (or $M_a$ for all-active RA array). In Fig.~\ref{fig:5}, we note that MAMP RA with greedy RAPS outperforms HSC FPA by about 34\% thanks to RA's position adjustments and full signal control (amplitude/phase). However, for $M=16$, HFC FPA surpasses MAMP RA with greedy/TS RAPS by about 3.8\%, due to the small number of ports/RA and large number of antennas.

As we notice in Fig.~\ref{fig:6}, for $M_a = 3$ and $M_p = 6$ ($M=9$), SR increases with $N=4\!:\!4\!:\!12$ and for $N\geq 9$, both  MAMP and all-active RA array using greedy RAPS outperform HFC and HSC FPA arrays, respectively, thanks to the enhanced spatial sampling. Likewise, with $N=10$ and $M_a = 2$, varying $M_p = [2; 4; 8; 12]$ (corresponding to $M=[4; 6; 10; 14]$) reveals that MAMP RA with greedy RAPS achieves about 57\% higher SR than HFC FPA until $M_p > 10$, where the array gain of the latter starts outweighing the spatial sampling benefits of the former, as shown in Fig.~\ref{fig:7}. This result highlights dynamic RA positioning benefits. SR improves with $M_p$ in both architectures, though MAMP RA experiences diminishing returns due to the constrained passive RA port selection in greedy RAPS and the increased inter-element spacing (reduced coupling). WF precoding outperforms ZF, as expected. Finally, Fig.~\ref{fig:8} shows that all RAPS algorithms converge quickly ($\sim 5-15$ iterations) and efficiently handle up to  $10^{24}$ combinations.

\begin{figure*}[!t]
\begin{minipage}[b]{.324\linewidth}
	\centering 
	\includegraphics[width=\linewidth]{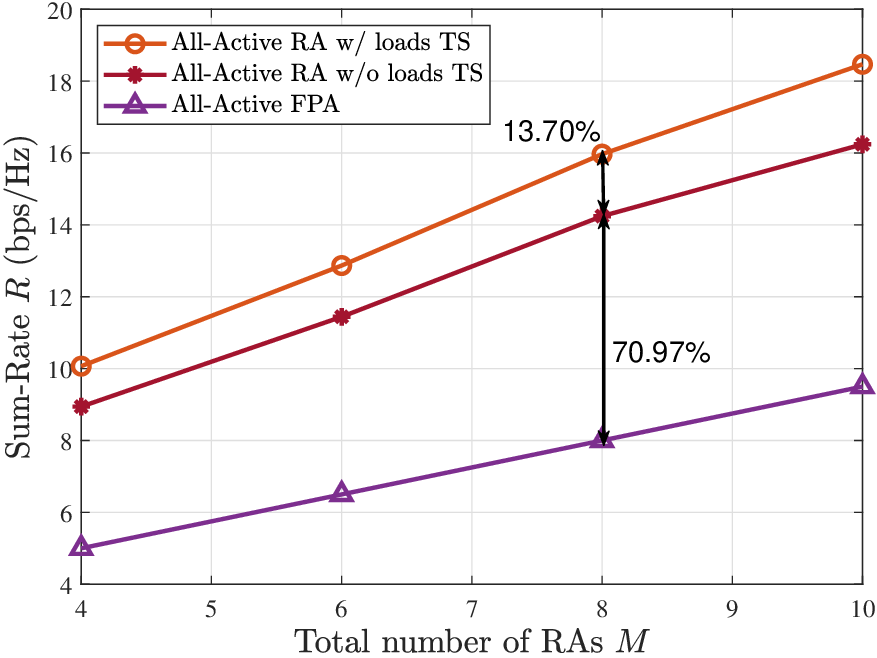}
	\caption{SR vs. number of RAs for the all-active array.}
	\label{fig:10}
\end{minipage}
	\hspace{1mm}
\begin{minipage}[b]{.324\linewidth}
	\centering 
	\includegraphics[width=\linewidth]{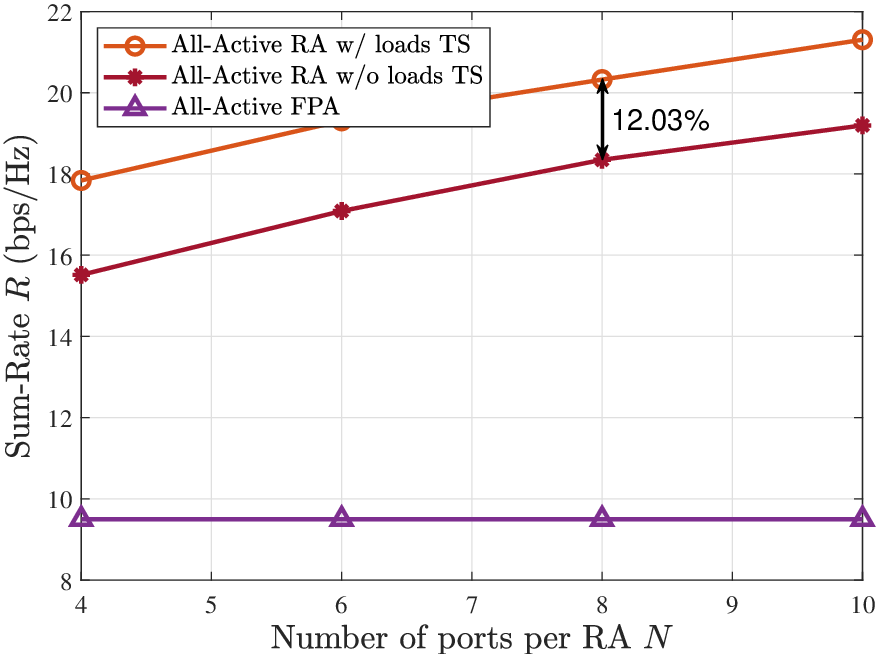}
	\caption{SR vs. number of ports for the all-active RA array.}
	\label{fig:11}
\end{minipage}
	\hspace{1mm}
\begin{minipage}[b]{.324\linewidth}
	\centering 
	\includegraphics[width=\linewidth]{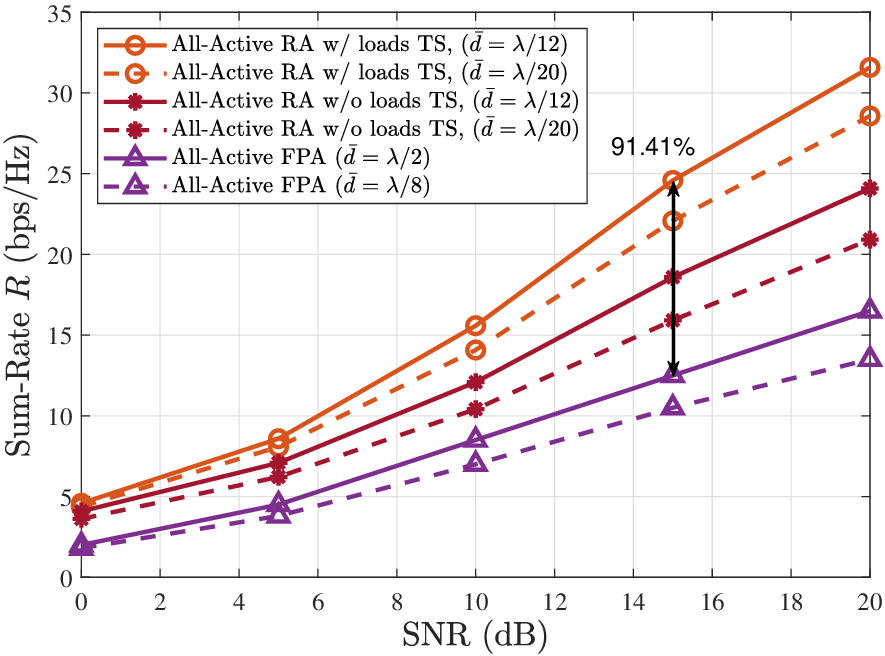}
	\caption{SR vs. SNR for all-active RA array with different antenna spacing.}
	\label{fig:12}
\end{minipage}
\end{figure*}

\begin{figure}[t!]
\centering 
\includegraphics[width=\columnwidth]{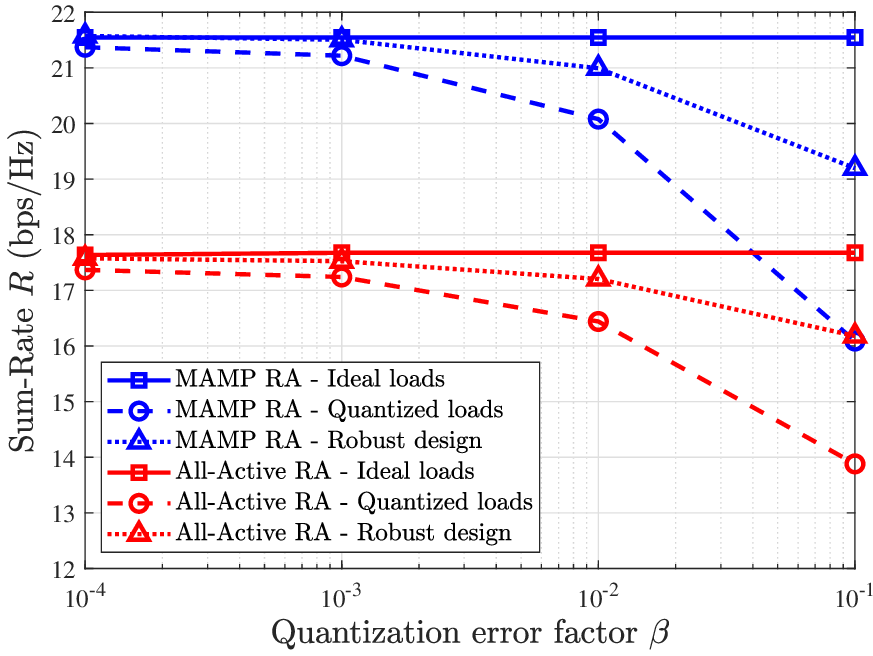}
\caption{SR vs. quantization error factor.}
\label{fig:13}
\end{figure}

For completeness, Figs.~\ref{fig:9a}--\ref{fig:9c} evaluate the performance of a MAMP RA array operating at 7 GHz against that of a respective MAMP FPA array utilizing the design presented in~\cite{Heath2025}, as well as the performance of an HSC transceiver with $M$ FPAs and $M_a$ RF chains and a fully-digital (FD) transceiver with $M_a$ FPAs. We investigate the achieved SR, energy efficiency ($EE = R/P$), and power consumption $P$. The latter varies by architecture: for the MAMP FPA/RA array, $P = P_{\max}+M_a P_{\mathrm{RFC}} + M_pP_{\mathrm{var/load}}$; for the FD transceiver, $P = P_{\max}+M_a P_{\mathrm{RFC}}$; and for the HSC design, $P = \epsilon P_{\max} + M_a P_{\mathrm{RFC}} + M_a(M_p+1)P_{\mathrm{PS}}$, where $P_{\mathrm{RFC}}=240$ mW, $P_{\mathrm{var}}\approx 0$, $P_{\mathrm{load}}=10$ mW, and $P_{\mathrm{PS}}=30$ mW respectively represent the power consumption of each RF chain, varactor, complex impedance load, and phase shifter, while $\epsilon=2.3$ dB accounts for the insertion losses of splitters and phase shifters~\cite{Heath2025}. For $M=8$, $M_a=M_p=4$, and $N=6$, we notice that the RA structure significantly outperforms all designs in SR and EE, highlighting the benefits of antenna repositioning and parasitic elements, with BPSO providing around 9\% higher gains than greedy RAPS (which is, however, 1.5-2.5 times faster). The EE improvements stem from both SR enhancement and power savings, attributed to the use of fewer RF chains and the lack of other components with high power consumption or insertion losses, as confirmed in Fig.~\ref{fig:9c}. The FPA variant of the MAMP design consumes about 3\% less power than the RA one, thanks to the use of varactors.

\subsection{All-Active RA Array With Tunable Loads}\label{subsec:6.2}

In this section, we evaluate the performance of the proposed all-active RA array with tunable loads in terms of the achievable SR. Two benchmarks are considered; an all-active RA array that does not incorporate analog loads, and an all-active FPA array with $\bar{d} = \lambda/2$. To ensure a fair comparison with the FPA array, the results are generated for the same number of antennas $M$, and only the number of ports per antenna $N$ are varied for the proposed all-active RA architecture. Since we compared the different RAPS variants in Sec.~\ref{subsec:6.1} and they perform close to each other, here we assume the use of TS only. Furthermore, we consider only the WF linear precoder, which provided the best performance. Unless otherwise stated, we set $f_c = 3$ GHz, $\bar{d} = \lambda/4$, $K=2$, $N=4$, $N_r = 50$, $\tilde{\phi} = \pi/8$ and a signal-to-noise-ratio (SNR) of 20 dB.

In Fig. \ref{fig:10}, we vary the number of RAs as $M = 4:2:10$. We notice that the use of analog loads improves performance by at least 13\%. This is attributed to the exploitation of the solution space without the limiting antenna spacing constraints. Furthermore, antenna repositioning ability adds another 71\% gain over the FPA variant. The performance of all schemes improves with $M$. We note that, compared with the MAMP RA array, the all-active RA array demonstrates a $23.1\%$ gain in performance for $M=8$ RAs (refer to Fig. \ref{fig:4}). We observe similar performance gaps in Fig. \ref{fig:11}, where we fix $M = 4$ and vary the number of ports per RA. We also note that the performance improves with $N$ for all RA-equipped arrays.

In Fig. \ref{fig:12}, we fix $N=4$ and vary the SNR as $\mathrm{SNR} = 0:5:20$ dB. We also assume either $\bar{d} = \lambda/12$ or $\bar{d} = \lambda/20$ for the RA arrays and $\bar{d} = \lambda/2$ or $\bar{d} = \lambda/8$ for the FPA one. We note that the SR increases with the SNR, as expected. We also observe that the loads-equipped all-active RA array performs much better than its counterpart without loads, as it is more robust against small antenna spacing (e.g., $\bar{d}=\lambda/12$). Evidently, after some point, there is a noticeable performance degradation ($\bar{d}=\lambda/20$), however, the variant with loads performs better than the one without. Moreover, we remark that the RA array without loads is more robust to antenna spacing shrinking than the FPA array.

\subsection{Robust Design with Quantized Loads}\label{subsec:6.3}

\begin{figure*}[!t]
\normalsize
\setcounter{mytempeqncnt}{\value{equation}}
\setcounter{equation}{41}
\begin{subequations}\label{eq:app1}
\begin{alignat}{2}
R_{m,m}= & \frac{\eta}{2 \pi}\left\{\ln (u)+\operatorname{Ci}(u)+0.5 \sin(u)\left[\operatorname{Si}(2u)-2 \operatorname{Si}(u)\right]+0.5 \cos (u)\left[\gamma +\ln (u/2)+\operatorname{Ci}(2u)-2 \operatorname{Ci}(u)\right]\right\}+\frac{\gamma \eta}{2 \pi}, \label{eq:app1a} \\
X_{m,m}= & \frac{\eta}{4 \pi}\left\{2 \operatorname{Si}(u)+\cos (u)\left[2 \operatorname{Si}(u)-\operatorname{Si}(2u)\right]-\sin (u) \left[2 \operatorname{Ci}(u)-\operatorname{Ci}(2u)-\operatorname{Ci}\left(2 \kappa a^2 / l\right)\right]\right\}. \label{eq:app1b}
\end{alignat}
\end{subequations}
\setcounter{equation}{\value{mytempeqncnt}}
\hrulefill
\end{figure*}

In this section, we evaluate the proposed robust designs when the tunable loads are quantized. In Fig.~\ref{fig:13}, we define the quantization error variance as $\epsilon = \beta \bar{z}$, where $\bar{z} = \frac{1}{M_a}\sum_{m\in\mathcal{M}_a}\left|z_m\right|^2$ (MAMP array) or $\bar{z} = \frac{1}{M}\sum_{m\in\mathcal{M}}\left|z_m\right|^2$ (all-active RA array), and vary $\beta$ as $\beta\in\left\{10^{-4},10^{-3},10^{-2},10^{-1}\right\}$. We note that, for $\beta=10^{-1}$, the MAMP array exhibits a performance loss of $25\%$ with a recovery of $57\%$ for the robust design. As for the all-active array, the performance loss is about $22\%$ with a recovery of $61\%$ for the robust case, highlighting the effectiveness of the robust designs against load quantization errors.

\section{Summary, Conclusions, and Future Work}\label{sec:7}
In this work, we introduced novel RA arrays (MAMP and all-active arrays) that utilize tunable analog loads to either mitigate or exploit the effect of mutual coupling, thus unlocking the full solution space and enhancing performance. We present various low-complexity RAPS algorithms that efficiently handle a vast number of array configurations. Numerical simulations unveil the significant performance gains of the proposed designs, as the MAMP and the all-active RA arrays outperform their FPA counterparts by 23$\%$ and $70\%$, respectively. Moreover, the proposed architectures showcase the effectiveness of the robust designs against load quantization errors, leading to a performance recovery of at least 57$\%$.

In the future, we plan to extend this work by considering realistic matching network constraints and investigating the feasibility of the proposed transceivers in emerging applications such as radar sensing or wireless power transfer.

\appendices

\renewcommand{\thesectiondis}[2]{\Alph{section}:}
\section{Self and Mutual Impedances}\label{App:A}
We denote as $\operatorname{Ci}(x) =- \int_x^{\infty} \frac{\cos(t)}{t}{\rm{d}}t$ and $\operatorname{Si}(x) = \int_{0}^x \frac{\sin(t)}{t}{\rm{d}}t$ the cosine and sine integral functions, respectively, $\kappa = 2\pi/\lambda$ the wavenumber, $\eta = 120\pi$ the intrinsic free-space impedance, and $\gamma = 0.5772$ Euler's constant, while $l$ represents the wavelength normalized RA's length and $a$ refers to each RA's radius, under a dipole approximation in both cases. Moreover, we define $\mu_0\!\triangleq\!\kappa d_{m,m^{\prime}}$, $\mu_1\!\triangleq\!\kappa\left(\sqrt{d_{m,m^{\prime}}^2+l^2}+l\right)$, and $\mu_2\!\triangleq\!\kappa\left(\sqrt{d_{m,m^{\prime}}^2+l^2}-l\right)$, where $d_{m,m^{\prime}}=\left|t_{m^{\prime}}-t_m\right|$ is the wavelength-normalized distance between the $m$-th and $m^{\prime}$-th RA, as well as $u\triangleq\kappa l\lambda$. Then, the self-impedances (diagonal entries of $\mathbf{Z}$) are computed as shown in Eq.~\eqref{eq:app1} on top of the page. Likewise, the mutual impedances (off-diagonal entries of $\mathbf{Z}$) are given by
\addtocounter{equation}{1}
\begin{subequations}\label{eq:app2}
\begin{alignat}{2}
R_{m,m^{\prime}} & =\frac{\eta}{4 \pi}\left[2 \operatorname{Ci}\left(\mu_0\right)-\operatorname{Ci}\left(\mu_1\right)-\operatorname{Ci}\left(\mu_2\right)\right], \label{eq:app2a} \\[2mm]
X_{m,m^{\prime}} & =\frac{\eta}{4 \pi}\left[2 \operatorname{Si}\left(\mu_0\right)-\operatorname{Si}\left(\mu_1\right)-\operatorname{Si}\left(\mu_2\right)\right]. \label{eq:app2b}
\end{alignat}
\end{subequations}

\bibliographystyle{IEEEtran}
\bibliography{refs2}

\end{document}